\newcommand*{\ATLASLATEXPATH}{}
\documentclass[cernpreprint, texlive=2016, txfonts, UKenglish]{\ATLASLATEXPATH atlasdoc}
\usepackage{\ATLASLATEXPATH atlaspackage}
\usepackage[backref=false]{\ATLASLATEXPATH atlasbiblatex}
\usepackage[process]{\ATLASLATEXPATH atlasphysics}
 
\graphicspath{{logos/}{figures/}}
 
\usepackage{ANA-STDM-2017-13-PAPER-defs}
\usepackage{booktabs}
 
\AtlasTitle{Measurement of the cross-section and charge asymmetry of $W$ bosons produced in proton--proton collisions at $\sqrt{s}=8~\text{Te\kern -0.1em V}$ with the ATLAS detector}
 
\AtlasAbstract{
This paper presents measurements of the $W^+ \rightarrow \mu^+\nu$\xspace and $W^- \rightarrow \mu^-\nu$\xspace cross-sections and the associated charge asymmetry as a function of the absolute pseudorapidity of the decay muon. The data were collected in proton--proton collisions at a centre-of-mass energy of 8~$\text{Te\kern -0.1em V}$ with the ATLAS experiment at the LHC and correspond to a total integrated luminosity of $20.2~\mbox{fb$^{-1}$}$. The precision of the cross-section measurements varies between 0.8\% to 1.5\% as a function of the pseudorapidity, excluding the 1.9\% uncertainty on the integrated luminosity. The charge asymmetry is measured with an uncertainty between 0.002 and 0.003. The results are compared with predictions based on next-to-next-to-leading-order calculations with various parton distribution functions and have the sensitivity to discriminate between them.
}
 
\author{The ATLAS Collaboration}
 
\AtlasRefCode{STDM-2017-13}
\PreprintIdNumber{CERN-EP-2019-053}
 
\AtlasJournal{Eur. Phys. J. C }
\AtlasJournalRef{Eur. Phys. J. C 79 (2019) 760}
\AtlasCoverEgroupEditors{atlas-STDM-2017-13-editors@cern.ch}
\AtlasDOI{10.1140/epjc/s10052-019-7199-0}
 
\AtlasCoverEgroupEdBoard{atlas-STDM-2017-13-editorial-board@cern.ch}
 
\hypersetup{pdftitle={ATLAS document},pdfauthor={The ATLAS Collaboration}}
 
\begin{document}
 
\maketitle
\newpage
\newpage
\clearpage
 
\section{Introduction}
 
Measurements of the \Wplus{} and \Wminus{} boson cross-sections in hadron collisions are a sensitive probe of quantum chromodynamics (QCD).
High-precision predictions at next-to-next-to-leading-order (NNLO) accuracy in QCD are available to compare with data.
Of particular interest is the ability of such measurements to discriminate between different parton distribution functions (PDFs)~\cite{Dulat:2015mca,STDM-2012-20,Abramowicz:2015mha,Ball:2017nwa,Butterworth:2015oua,Harland-Lang:2014zoa,Alekhin:2017kpj},
because the \Wboson{} boson rapidity\footnote{ATLAS uses a right-handed
coordinate system with its origin at the nominal interaction point
in the centre of the detector and the $z$-axis coinciding with the axis of the beam pipe.
The $x$-axis points from the interaction point to the centre of the LHC ring, and the $y$-axis points upward.
Polar coordinates ($r$, $\phi$) are used in the transverse plane, $\phi$ being the azimuthal angle around the beam pipe.
The pseudorapidity is defined in terms of the polar angle $\theta$ as $\eta=-\ln\tan(\theta/2)$.
The rapidity $y$ of a system is defined in terms of its energy $E$
and its longitudinal momentum $p_z$ as $y = (1/2)\ln[(E+p_z)/(E-p_z)]$.
Angular separations between particles or reconstructed objects are measured in $\eta$--$\phi$
space using $\Delta R = \sqrt{(\Delta \eta)^2 + (\Delta \phi)^2}$.}
$y$ is strongly correlated with the initial-state parton momentum fractions $x$.
In high-energy proton--proton collisions, the main production mechanism of single \Wboson{} bosons is a valence quark annihilating with a sea antiquark.
The \Wboson{} bosons are preferentially produced with a boost in the direction of the incoming valence quark,
as the quark is more likely to be at a higher $x$ than the corresponding antiquark.
Since the PDFs of $u$ and $d$ quarks in the proton differ
(largely due to there being two valence $u$ quarks and one valence $d$ quark),
there is a production asymmetry between \Wplus{} and \Wminus{} bosons (referred to in this paper as the \Wboson{} boson charge asymmetry),
which also varies as a function of rapidity.
The boson rapidity cannot be determined unambiguously in leptonic decays of the \Wboson{} boson because the decay neutrino passes through the detector unobserved.
The charge asymmetry can instead be measured as a function of the decay lepton's pseudorapidity $\eta_{\ell}$,
which is strongly correlated with the \Wboson{} boson rapidity.

The \Wboson{} boson charge asymmetry was measured
in proton--antiproton collisions by the CDF and D0 collaborations~\cite{Aaltonen:2009ta,Abazov:2013rja2,D0:2014kma}.
It was also measured, along with the individual cross-sections, in proton--proton collisions at the LHC by the ATLAS Collaboration at
centre-of-mass energies of \sqs\ = 5~\TeV~\cite{HION-2018-02} and  7~\TeV~\cite{STDM-2012-20},
by the CMS Collaboration at \sqs\ = 7 and 8~\TeV{}~\cite{CMS-SMP-12-001,CMS-SMP-12-021,CMS-SMP-14-022},
and by the LHCb Collaboration at \sqs\ = 7 and 8~\TeV{}~\cite{Aaij:2014wba,Aaij:2015zlq,Aaij:2016qqz}.

This paper presents measurements of the integrated fiducial cross-sections for \Wpmn{} and \Wmmn{},
as well as the differential cross-sections,
$\mathrm{d}\sigma_{W_{\mu^{+}}}/\mathrm{d}\eta_{\mu}$ and $\mathrm{d}\sigma_{W_{\mu^{-}}}/\mathrm{d}\eta_{\mu}$,
as a function of $|\eta_{\mu}|$, where $\eta_{\mu}$ is the pseudorapidity of the decay muon.
The data used were collected in proton--proton collisions at a centre-of-mass energy of $\sqrt{s}=8$~\TeV{} with the ATLAS experiment at the LHC and correspond to a total integrated luminosity of $20.2~\ifb$~\cite{DAPR-2013-01}.
The muon decay channel (\Wmn{}) is particularly well suited for this measurement due to good lepton identification and small contributions from background processes.
In addition, a measurement of the \Wboson{} boson charge asymmetry $A_{\mu}$ is presented, also as a function of $|\eta_{\mu}|$.
The asymmetry is defined in terms of the $W^+$ and $W^-$ differential cross-sections as
 
\begin{equation}
A_{\mu} = \frac{\mathrm{d}\sigma_{W_{\mu^{+}}}/\mathrm{d}\eta_{\mu} - \mathrm{d}\sigma_{W_{\mu^{-}}}/\mathrm{d}\eta_{\mu}}{\mathrm{d}\sigma_{W_{\mu^{+}}}/\mathrm{d}\eta_{\mu} + \mathrm{d}\sigma_{W_{\mu^{-}}}/\mathrm{d}\eta_{\mu}}\,\,.
\label{eq:muonchargeasymmetry}
\end{equation}
 
The measurements are performed in a fiducial phase space,
which is defined by the kinematics and geometrical acceptance of the muon.
All measurements are compared with predictions from a calculation performed at NNLO accuracy using the DYNNLO program~\cite{Catani:2009sm}.
The DYNNLO predictions are produced with six different PDF sets.

 
\section{The ATLAS detector}
 
\newcommand{\AtlasCoordFootnote}{
ATLAS uses a right-handed coordinate system with its origin at the nominal interaction point (IP)
in the centre of the detector and the $z$-axis along the beam pipe.
The $x$-axis points from the IP to the centre of the LHC ring,
and the $y$-axis points upwards.
Cylindrical coordinates $(r,\phi)$ are used in the transverse plane,
$\phi$ being the azimuthal angle around the $z$-axis.
The pseudorapidity is defined in terms of the polar angle $\theta$ as $\eta = -\ln \tan(\theta/2)$.
Angular distance is measured in units of $\Delta R \equiv \sqrt{(\Delta\eta)^{2} + (\Delta\phi)^{2}}$.}
 
The ATLAS detector~\cite{PERF-2007-01} at the LHC covers nearly the entire solid angle around the collision point.
It consists of an inner tracking detector (ID) surrounded by a thin superconducting solenoid, electromagnetic and hadronic calorimeters,
and a muon spectrometer (MS) incorporating three large superconducting toroid magnets.
The ID is immersed in a \SI{2}{\tesla} axial magnetic field
and provides charged-particle tracking in the range $|\eta| < 2.5$.
A high-granularity silicon pixel detector typically provides three measurements per track and
is followed by a silicon microstrip tracker, which usually provides four three-dimensional measurement points per track.
These silicon detectors are complemented by a transition radiation tracker,
which enables radially extended track reconstruction up to $|\eta| = 2.0$.
 
The calorimeter system covers the pseudorapidity range $|\eta| < 4.9$.
Electromagnetic calorimetry is provided by barrel and endcap high-granularity lead/liquid-argon (LAr) sampling calorimeters
in the $|\eta|< 3.2$ region.
A thin LAr presampler, covering $|\eta| < 1.8$, corrects for energy loss in the material upstream of the calorimeters.
Hadronic calorimetry is provided in the $|\eta| < 1.7$ region by the steel/scintillator tile calorimeter,
segmented into three barrel structures, and in the endcap region by two copper/LAr calorimeters.
The forward regions $3.1<|\eta|<4.9$ are covered by copper/LAr and tungsten/LAr calorimeter modules
optimised for electromagnetic and hadronic measurements, respectively.
 
The MS has separate trigger and
precision tracking chambers measuring the deflection of muons in a magnetic field generated by superconducting air-core toroids.
The precision chamber system covers the region $|\eta| < 2.7$ with three layers of monitored drift tubes,
complemented by cathode-strip chambers in the forward regions $2.0 < |\eta| < 2.7$, where the background is highest.
There is a transition between the barrel and endcap muon detectors around $|\eta| = 1.05$.
The muon trigger system covers the range of $|\eta| < 2.4$ with resistive-plate chambers in the barrel and thin-gap chambers in the endcap regions.
 
A three-level trigger system~\cite{PERF-2011-02,Aad:2014sca} selected candidate events in 2012.
The level-1 trigger was implemented in hardware and used a subset of detector information
to reduce the event rate to a design value of at most \SI{75}{\kHz}.
This was followed by two software-based trigger levels which together reduced the event rate to about \SI{400}{\Hz}.
 
\section{Analysis methodology}
 
\subsection{Description of the measurements}
 
The integrated cross-sections for \Wpmn{} and \Wmmn{} production
are measured in a fiducial phase space defined at the particle level by requiring the muon transverse momentum
\ptm{} to be greater than 25~\GeV{} and the neutrino transverse momentum \ptn{}
to be greater than 25~\GeV{}.
The absolute muon pseudorapidity is required to be less than 2.4.
The \Wboson{} boson transverse mass is
\begin{equation}
\mt = \sqrt{ 2 \ptm \ptn \left( 1- \cos{\left( \phi^{\mu}-\phi^{\nu}  \right)} \right) }\, ,
\label{eq:mt}
\end{equation}
where $\phi^{\mu}$ and $\phi^{\nu}$ are
the azimuthal angles of the muon and neutrino, respectively.
For this analysis, \mt\ must be at least 40~\GeV, both at reconstruction and particle level. The requirements on fiducial quantities are defined before the emission of final-state photon radiation (i.e.\ at the `Born level').
 
The differential cross-sections and charge asymmetry are measured in the
same fiducial phase space as for the integrated measurement.
These are  measured in 11 bins of absolute muon pseudorapidity between 0 and 2.4
with bin edges at 0, 0.21, 0.42, 0.63, 0.84, 1.05, 1.37, 1.52, 1.74, 1.95, 2.18, and 2.4.
The bin edges are identical to those used in the ATLAS 7~\TeV{} measurement~\cite{STDM-2012-20}.

\subsection{Data and simulated event samples}
\label{sec:dataandmc}
 
The data for this analysis comprise the entire ATLAS $\sqs=8$~\TeV{} data set recorded
between April and December 2012,
corresponding to an integrated luminosity of $20.2~\ifb$.
The average number of proton--proton interactions per bunch crossing $\langle\mu\rangle$ was 20.7.
Only events recorded with stable beams and the detector operating well are selected.
The relative uncertainty of the LHC proton beam energy of $\pm0.1\%$~\cite{Todesco:2017nnk} has no significant effect on the results.
 
Events from Monte Carlo (MC) simulations, including simulation of the ATLAS detector, are used for the background estimation and
to correct the measured data for detector acceptance, efficiency, and resolution effects.
 
The \Wmn{} signal process was simulated using \POWHEGBOX{}~\cite{Alioli:2008gx,Alioli:2010xd} at next-to-leading order (NLO) in perturbative QCD
using the \texttt{CT10} set of PDFs~\cite{Lai:2010vv} and interfaced to
\Pythia{} 8.170~\cite{Sjostrand:2007gs} with the AU2 set of tuned parameters~\cite{ATL-PHYS-PUB-2012-003}
to simulate the parton shower, hadronisation, and underlying event and to \Photos{}~\cite{Golonka:2005pn} to simulate final-state photon radiation (FSR).
This is referred to as \Powheg{}+\Pythia{}8 in this paper.
An alternative signal sample was simulated using NLO~\Sherpa{} 1.4~\cite{Gleisberg:2008ta} and the \texttt{CT10} PDF set to cross-check the results obtained using~\Powheg{}+\Pythia{}8 and to evaluate systematic uncertainties in the signal modelling.
 
\Powheg{}+\Pythia{}8 with the \texttt{CT10} PDF set was used to simulate the background processes \Wtaunu{} and~\Zmm{} (with the AU2 tune set)
and \Powheg{}+\Pythia{}6 \cite{Frixione:2007nw, Sjostrand:2006za} also with the \texttt{CT10} PDF set was used to simulate the \ttbar{} process (with the P2011C tune set \cite{ATL-PHYS-PUB-2014-021}).
The \Ztautau{} process was simulated using \Alpgen{}~\cite{Mangano:2002ea} with the \texttt{CTEQ6L1} PDF set~\cite{Pumplin:2002vw}
interfaced to \Herwig{}~\cite{Corcella:2000bw} to simulate the parton shower and
\Jimmy{}~\cite{Butterworth:1996zw} to model the underlying event.
The single-top process in the \textit{s}-channel and \textit{Wt}-channels was simulated with \MCatNLO{}~\cite{Frixione:2002ik} interfaced to \JIMMY{}.
The \textit{t}-channel was generated with \ACERMC{}~\cite{Kersevan:2004yg} interfaced to \Pythia{}.
The backgrounds from the diboson processes $WW$, $WZ$, and $ZZ$ were simulated using \Herwig{} at leading order with the \texttt{CTEQ6L1} PDF set.
In all cases, the \Geant4~\cite{Agostinelli:2002hh} program was used to simulate the passage of particles through the ATLAS detector \cite{Aad:2010ah}.
The multijet background is estimated using a data-driven technique,
described in Section~\ref{sec:Backgrounds}.
A $b\bar{b}\rightarrow\mu X$ MC sample, simulated using \Pythia{}8 with the AU2 tune set, is used to cross-check the data-driven estimation.
 
Differences in reconstruction, trigger, and isolation efficiencies for muons between MC simulations and data are
evaluated using a tag-and-probe method~\cite{PERF-2014-05}
and are corrected by reweighting the MC simulated events.
The muon reconstruction efficiencies are parameterised versus $\eta$ and $\phi$,
the muon isolation efficiencies versus $\eta$ and transverse momentum \pt{}, and the muon trigger efficiencies versus $\eta$, $\phi$, and \pt{}.
The reconstruction, trigger and isolation efficiencies are evaluated separately for positive and negative muons.
This separate evaluation is particularly necessary in the case of the trigger efficiencies,
which differ by up to 3\% (depending on $\eta$) between positive and negative muons,
much greater than the total uncertainty in the cross-section from other sources.
Corrections are also applied to MC events for the description of the muon momentum scales and resolution,
which are determined from fits to the observed mass distributions of~\Zmm{} candidates in data and MC simulations~\cite{PERF-2014-05}.
To correct for charge-dependent biases of the muon momentum scale due to residual misalignments in the ID and MS,
an additional momentum-dependent correction parameterised versus $\eta$ and $\phi$ is applied.
An associated uncertainty corresponding to the full size of the correction is included.
 
All simulated samples are normalised using their respective inclusive cross-sections at higher orders in perturbative QCD.
The \Wboson{} and \Zboson{} predictions are scaled to the NNLO calculation obtained with DYNNLO v1.5~\cite{Catani:2009sm,Catani:2007vq} and the \texttt{MSTW2008} PDF set~\cite{Martin:2009iq}.
The production of top quarks is normalised using the prediction at NNLO+NNLL precision from the Top++2.0 program for \ttbar{}~\cite{ttbarXsec1,ttbarXsec2,ttbarXsec3,ttbarXsec4,ttbarXsec5,ttbarXsec6,ttbarXsec7}, to the calculations in Refs.~\cite{Kidonakis:2010tc,Kidonakis:2011wy,Kidonakis:2010ux} for single top quarks,
and for diboson production to the NLO calculations following the procedure described in Ref.~\cite{ATL-PHYS-PUB-2017-005}.
 
The effect of multiple interactions per bunch crossing (pile-up) is simulated by
overlaying minimum-bias MC events generated with~\Pythia8 (with the A2 tune set)~\cite{SOFT-2010-01}.
The simulated event samples are reweighted to describe the
distribution of the number of pile-up events in the data.
The MC simulations are also reweighted to better describe the distribution of the
longitudinal position of the primary proton--proton collision vertex~\cite{ATLAS-CONF-2010-069} in data.
 
\subsection{Event selection}
 
Candidate \Wmn{} events are selected with the requirement that at least one of two single-muon triggers are satisfied.
A high-threshold trigger requires muons to have $\pt>36$~\GeV{},
whilst a low-threshold trigger requires $\pt>24$~\GeV{}
alongside the requirement that the muon must be isolated from additional nearby tracks.
 
Muon candidates are reconstructed by combining tracks measured in both the ID and
the MS~\cite{PERF-2014-05}.
The \pt{} of the ID track is required to be greater than 25~\GeV{} and the absolute pseudorapidity to be less than 2.4.
Track quality requirements are imposed for muon identification and background suppression.
The transverse impact parameter significance is required to be less than 3 to ensure that the muon candidates originate from a primary proton--proton interaction vertex.
The muon candidates are also required to be isolated, satisfying \Imu~$<$~0.1, where
\Imu\ is the scalar sum of the \pt{} of
tracks within a cone of size $\Delta R = 0.4$ around the muon (excluding the muon track) divided by the \pt{} of the
muon.
Events are required to contain exactly one muon candidate satisfying the above criteria.
 
To reduce background contamination, in particular from multijet processes,
events are required to have missing transverse momentum \MET{} greater than 25~\GeV{}.
The \MET{} is reconstructed using energy depositions in the calorimeters and tracks reconstructed in the inner detector and muon spectrometer~\cite{PERF-2014-04}.
It is defined as the absolute value of the negative of the vectorial sum of the transverse momenta of reconstructed objects (e.g.\ electrons, muons, jets)
and tracks not associated with these objects.
These are labelled the hard and soft terms, respectively.
 
The \Wboson{} boson transverse mass \mt{} is required to be larger than 40~\GeV.
This variable is defined analogously to Eq.~(\ref{eq:mt}) with \ptn{} replaced by \MET{}
and $\phi^{\nu}$ replaced by the azimuthal angle related to the \MET{}.
 
\subsection{Estimation of backgrounds}
\label{sec:Backgrounds}
 
The backgrounds from all sources other than multijet processes are estimated using the MC samples detailed in Section~\ref{sec:dataandmc}.
The \Zmm{} process with one of the muons not identified
contributes between 1\% and 8\% of selected data events, depending on the value of $\modetam$.
This is the largest background for $\modetam\gtrsim 1.4$.
The contribution from \Wtaunu{} where the $\tau$-lepton decays into a muon
is 2\% of the selected data events and approximately constant as a function of \modetam{}.
The backgrounds from \Ztautau{}, \ttbar{}, and the diboson processes $WW$, $WZ$, and $ZZ$
each amount to less than 0.3\% of the selected data.
 
Multijet processes contribute between 2\% and 4\% of the selected data
and are the largest sources of background for $\modetam\lesssim 1.4$.
The number and properties of the background events arising from multijet
processes are estimated using a data-driven template-fit technique, similar to that used in Ref.~\cite{STDM-2014-18}.
A multijet-dominated sample is obtained from data by selecting events passing the nominal selection except that the muon fails the isolation criterion.
For this purpose, events satisfying a trigger without an isolation requirement are used.
Multijet templates are constructed from this sample, in a series of mutually exclusive slices in muon isolation, for each distribution of interest.
The residual contribution from signal and other background sources, estimated from MC simulations, is subtracted.
The normalisations of the multijet templates in the signal region are obtained by fitting the templates to data in
three discriminating variables: \MET{}, \mt{}, and the ratio $\ptm/\mt$.
The fits are performed in two phase-space regions (fitting regions) in which the selections on \MET{} and \mt{} are relaxed
in order to enrich the multijet background contribution.
A requirement that  the \Wboson{} transverse momentum be less than 30~\GeV\ is introduced in one
fitting region to remove a region poorly modelled by signal MC simulations.
The normalisation of the multijet template is allowed to float in each of the fits and the total MC-based signal plus background is kept constant.
The multijet normalisation in the signal selection is extracted using each discriminating variable
in both fitting regions (a total of six estimates).
It is assumed that the normalisations of the templates in the fitting regions are the same as for the signal selection.
The fits described above are performed for each muon-isolation slice (i.e. six fits per slice).
The normalisation estimate extracted in each muon-isolation slice (for a particular discriminating variable and fitting region)
is used to linearly extrapolate to an isolation value of 0.05,
which is the average isolation of multijet events in the signal region as estimated using the $b\bar{b}\rightarrow\mu X$ MC simulation.

The central value of the multijet background is chosen to be the average of the extrapolated curves at $\Imu=0.05$,
and the spread gives one component of the systematic uncertainty,
which is estimated to be bin-to-bin correlated and 50\% charge correlated.
This uncertainty component includes the effects from the choice of the discriminating variable and the definition of the fitting regions.
Another component arises from the effect of the cross-section uncertainty on the signal simulation contribution in the fitting regions ($\pm 5\%$).
The fit is repeated varying this and the deviation from the nominal normalisation is taken as an uncertainty,
which is again treated as being charge and bin-to-bin correlated.
The fit also has a statistical uncertainty, which is treated as being uncorrelated between charges and bins.
This is largely due to the limited number of data events for the multijet templates passing the signal selection in each muon-isolation slice.
The size of the multijet systematic uncertainty is reported in Section~\ref{sec:Systematics}.
The above procedure is performed separately in each bin of muon pseudorapidity.
A summary of the most important backgrounds including systematic uncertainties is given in Table~\ref{tab:bkgratio}.
 
\begin{table}
\caption{The number of data events after event selection for \Wpmn{} and \Wmmn{} and the percentage of selected data that each of the three major backgrounds constitutes.}
\label{tab:bkgratio}
\centering
\begin{tabular}{ lcc }
\toprule
& \Wpmn & \Wmmn \\
\midrule
& \multicolumn{2}{c}{Number of events} \\
Data & \numprint{50390184} & \numprint{34877365} \\
\midrule
& \multicolumn{2}{c}{Percentage of data} \\
Multijet & 2.4\hphantom{0} $\pm$ 0.3\hphantom{0} & 3.1\hphantom{0} $\pm$ 0.3\hphantom{0} \\
\Wtaunu{} & 1.9\hphantom{0} $\pm$ 0.1\hphantom{0} & 2.0\hphantom{0} $\pm$ 0.1\hphantom{0} \\
\Zmm & 3.1\hphantom{0} $\pm$ 0.2\hphantom{0} & 4.0\hphantom{0} $\pm$ 0.2\hphantom{0} \\
Others & 0.62 $\pm$ 0.02 & 0.82 $\pm$ 0.03 \\
\bottomrule
\end{tabular}
\end{table}

Figure~\ref{fig:leptoneta} shows the muon $\eta$, muon \pt{}, and \Wboson{} transverse mass distributions of selected events with
positive muons (left) and with negative muons (right).
The data are compared with the sum of MC and data-based estimates for the signal and the backgrounds.
The predictions are normalised to the luminosity of the data,
after first normalising each MC cross-section to a corresponding higher-order prediction.
A normalisation shift between data and MC simulations of approximately 1\%  is observed for the positive muon plots.
This is covered by the uncertainty in the MC signal prediction due to the cross-section uncertainty. Otherwise, the combined prediction describes the data well and within the uncertainties.
 
\begin{figure}
\centering
\includegraphics[width=0.45\textwidth]{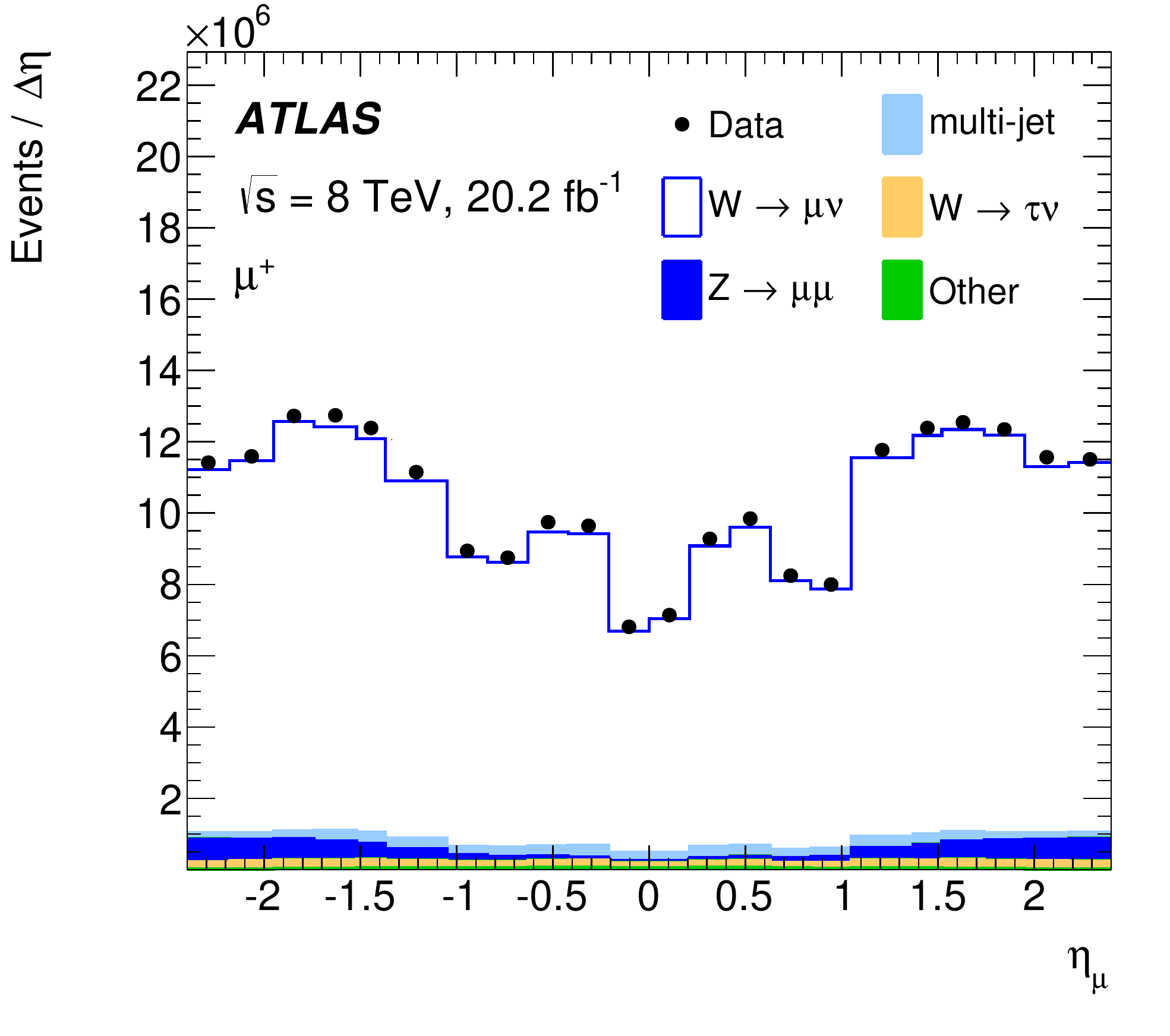}
\includegraphics[width=0.45\textwidth]{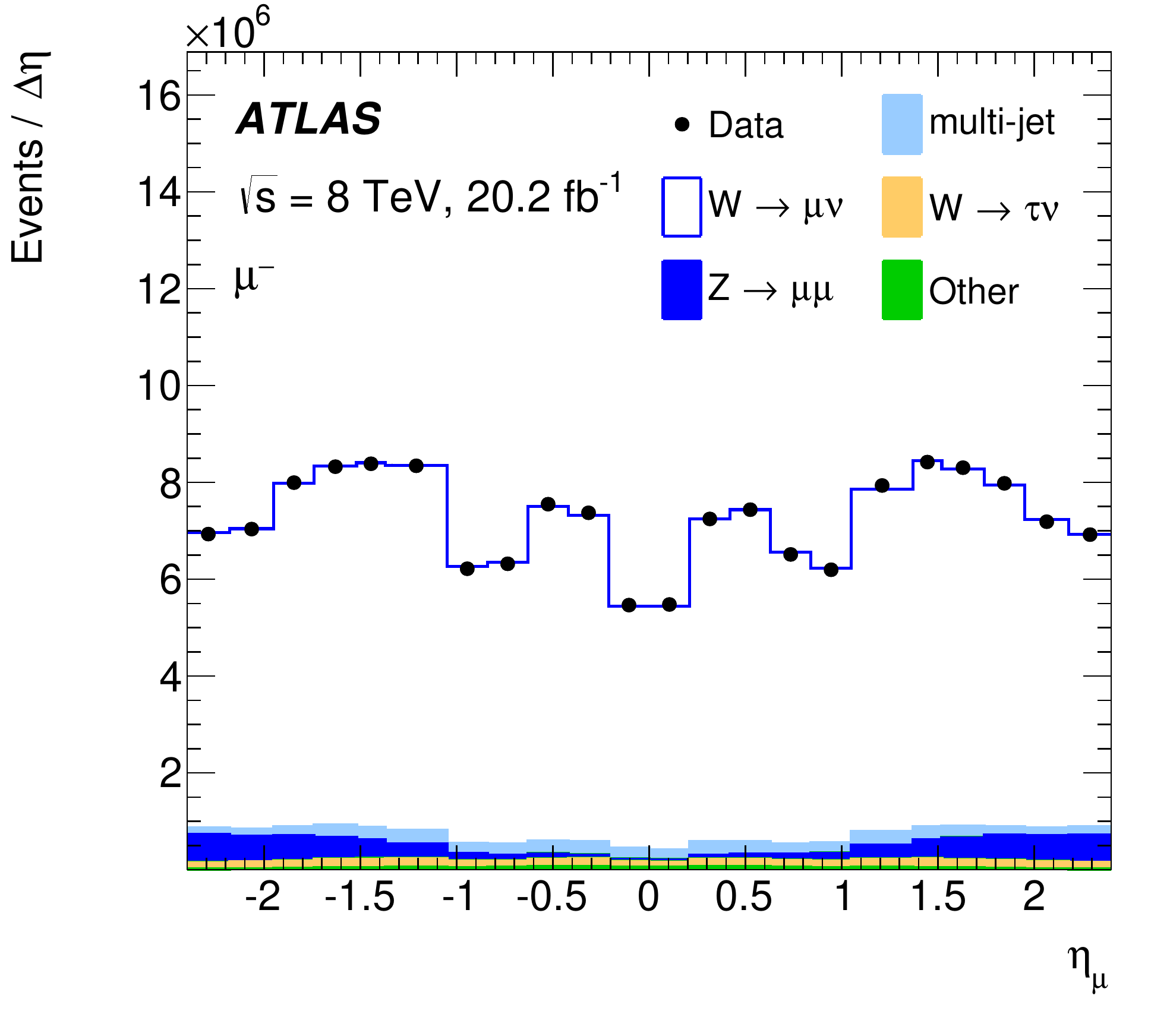}\\
 
\includegraphics[width=0.45\textwidth]{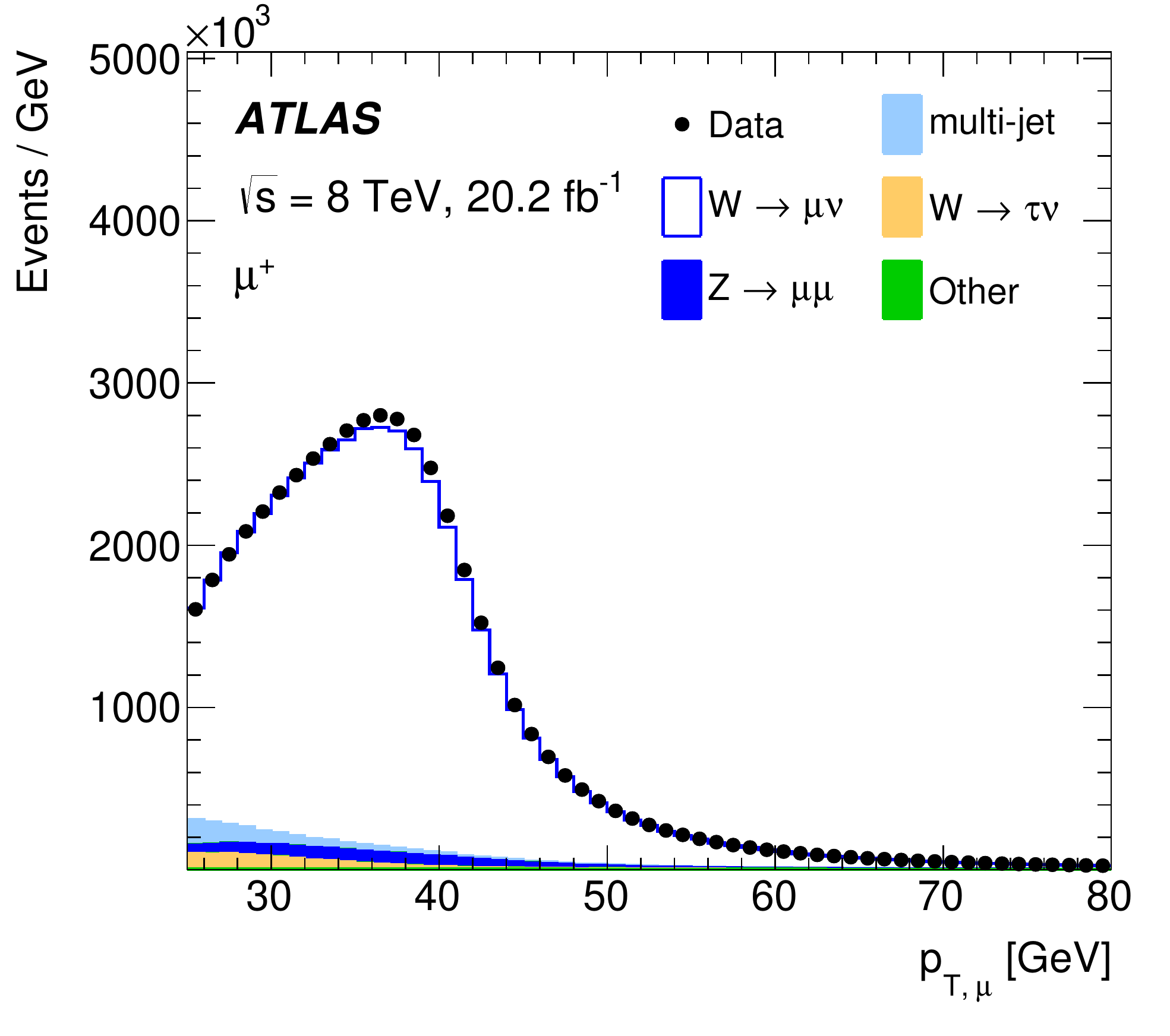}
\includegraphics[width=0.45\textwidth]{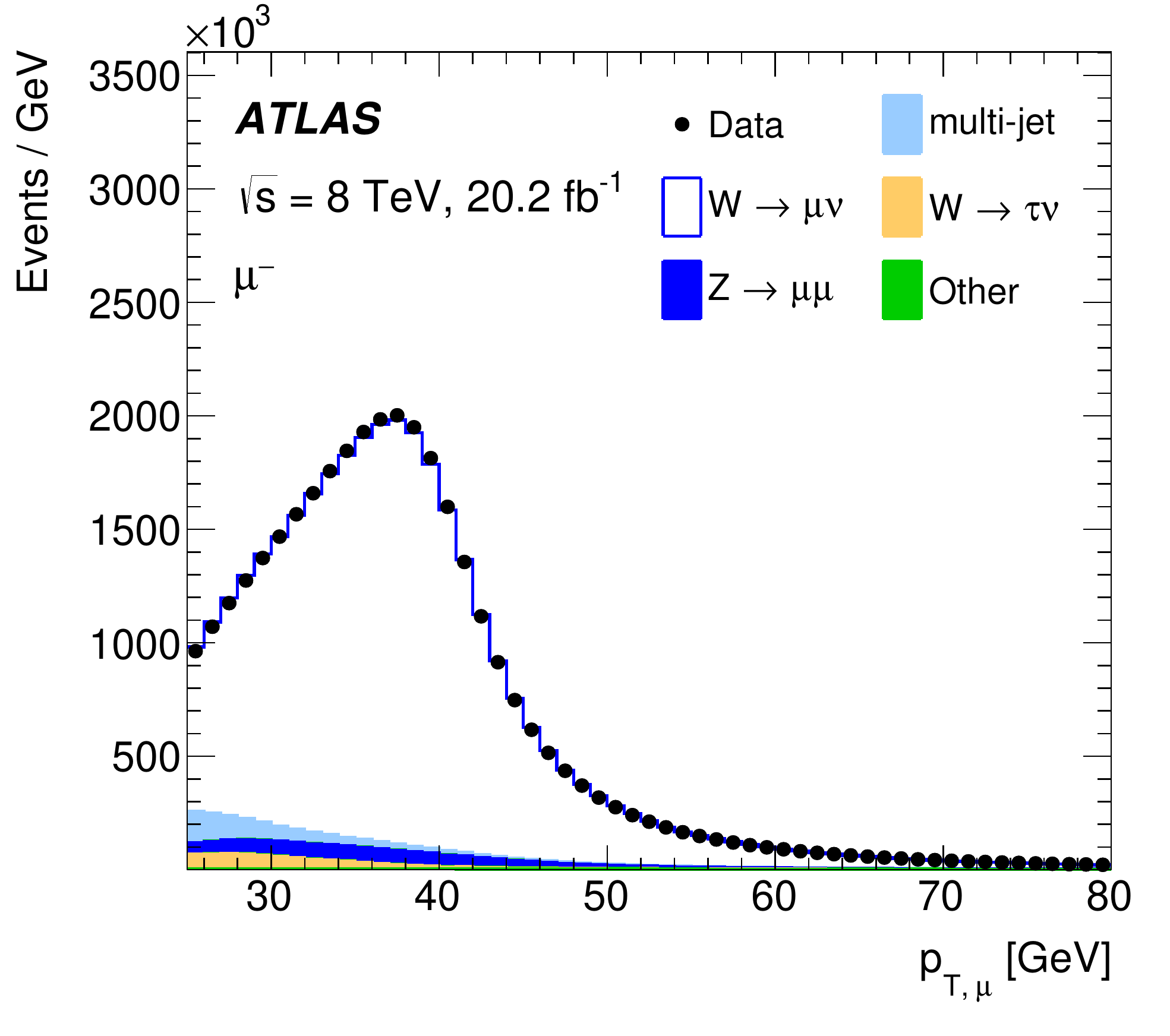}\\
 
\includegraphics[width=0.45\textwidth]{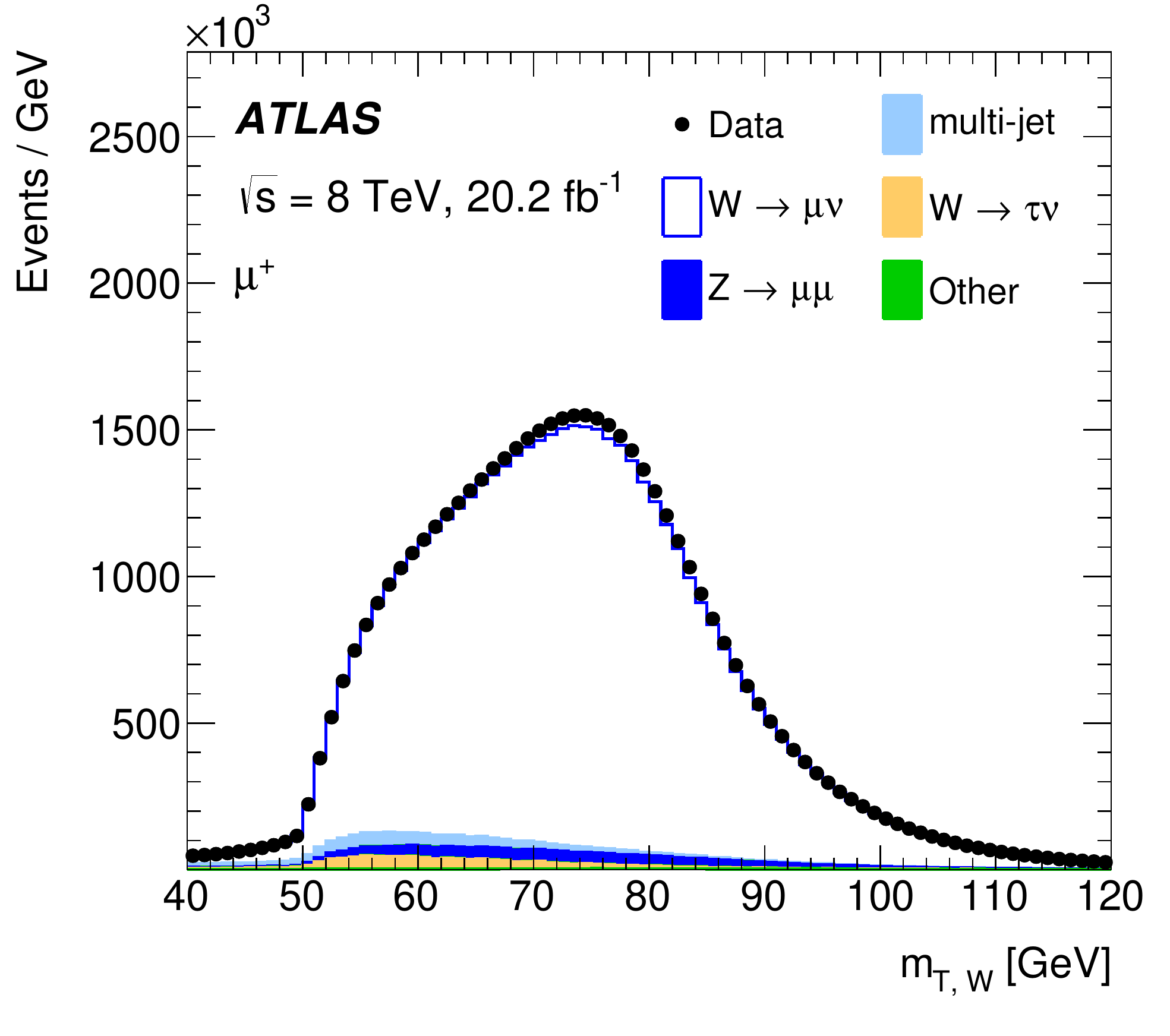}
\includegraphics[width=0.45\textwidth]{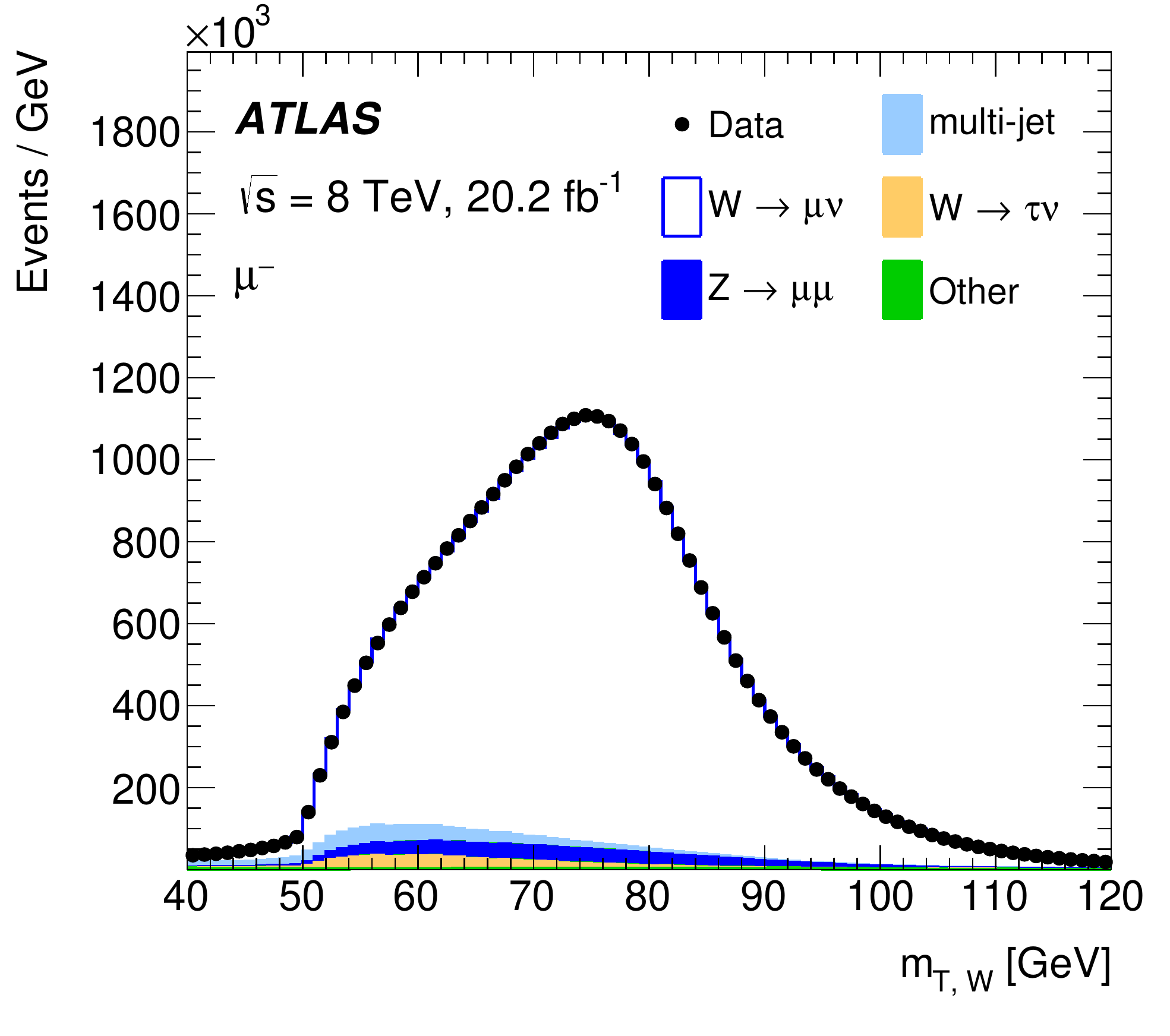}\\
 
\caption{The muon $\eta$ (top), muon \pT{} (centre), and \Wboson{} boson transverse mass (bottom) distributions of selected events with
positive muons (left) and negative muons (right).
The statistical uncertainties of the data points are smaller than the size of the markers.
}
\label{fig:leptoneta}
\end{figure}
 
\subsection{Obtaining the fiducial cross-sections}
 
The fiducial $W^{\pm}$ differential cross-sections in bin $i$ of pseudorapidity $(\mathrm{d}\sigma_{W_{\mu^{\pm}}}/\mathrm{d}\eta_{\mu})_i$ are obtained from
 
\begin{equation*}
\left(\frac{\mathrm{d}\sigma_{W_{\mu^{\pm}}}}{\mathrm{d}\eta_{\mu}}\right)_{i} = \frac{N_{\textrm{data},i} -  N_{\textrm{bkg},i}}{\Delta\eta_i \cdot C_{W^{\pm},i} \cdot \int\mathcal{L} \mathrm{d}t}\,,
\end{equation*}
 
where $N_{\textrm{data},i}$ is the number of selected candidate events in data,
$N_{\textrm{bkg},i}$ is the number of background events
estimated using the methods described in Section~\ref{sec:Backgrounds},
$\Delta\eta_i$ is the width of bin $i$, and $\int\mathcal{L} \mathrm{d}t$ is the integrated luminosity.
The results are provided as a function of absolute pseudorapidity, where $\Delta|\eta_i|= 2\cdot\Delta\eta_i$.
The term $C_{W^{\pm},i}$ is a factor (different for positive and negative channels) which corrects for the various detector inefficiencies and resolution effects.
This is estimated using \Wmn{} signal MC simulations and defined for each bin $i$ as the number of reconstructed events
satisfying the same selection criteria as data divided by the number of generated events in the fiducial phase space.
The charge asymmetry is then evaluated in each absolute pseudorapidity bin using Eq.~(\ref{eq:muonchargeasymmetry}).
A bin-by-bin correction method is used as the purity of each bin is 99\% or larger,
where the bin purity is defined as the ratio of events generated and reconstructed in a certain bin to all events reconstructed in that bin
(where all events are in the generator-level fiducial phase space).
 
The integrated fiducial cross-sections ($\sigma_{W_{\mu^{\pm}}}$) are obtained using one global correction factor, $C_{W^{{\pm}}}$,
(i.e.\ the total number of reconstructed events satisfying the same selection criteria as data divided by the total number of generated events in the integrated fiducial phase space).
The values are also obtained by summing the differential cross-sections as a function of \modetam{},
and the results are consistent.
Table~\ref{tab:cw-factors} lists the $C_{W^{\pm},i}$ values with their associated systematic uncertainties as a function of \modetam{}
and gives the integrated global correction factor $C_{W^{{\pm}}}$, each for \Wplus{} and \Wminus{}.
 
\begin{table}
\caption{The $C_{W^{\pm},i}$ values with their associated systematic uncertainties as a function of \modetam{} and the integrated global correction factor $C_{W^{{\pm}}}$, each for \Wplus{} and \Wminus{}.}
\label{tab:cw-factors}
\centering
\begin{tabular}{ c c c}
\toprule
\modetam{} & \Wpmn{} & \Wpmn{} \\
\midrule
0.00--0.21 & 0.508 $\pm$ 0.004 & 0.505 $\pm$ 0.004\\
0.21--0.42 & 0.684 $\pm$ 0.004 & 0.679 $\pm$ 0.004\\
0.42--0.63 & 0.702 $\pm$ 0.005 & 0.702 $\pm$ 0.005\\
0.63--0.84 & 0.611 $\pm$ 0.004 & 0.613 $\pm$ 0.005\\
0.84--1.05 & 0.603 $\pm$ 0.004 & 0.601 $\pm$ 0.005\\
1.05--1.37 & 0.795 $\pm$ 0.006 & 0.796 $\pm$ 0.007\\
1.37--1.52 & 0.848 $\pm$ 0.008 & 0.845 $\pm$ 0.007\\
1.52--1.74 & 0.861 $\pm$ 0.009 & 0.856 $\pm$ 0.007\\
1.74--1.95 & 0.856 $\pm$ 0.009 & 0.855 $\pm$ 0.008\\
1.95--2.18 & 0.792 $\pm$ 0.008 & 0.794 $\pm$ 0.009\\
2.18--2.40 & 0.802 $\pm$ 0.008 & 0.812 $\pm$ 0.011\\
\midrule
Integrated  & 0.736 $\pm$ 0.003 & 0.727 $\pm$ 0.003\\
\bottomrule
\end{tabular}
\end{table}
 
\section{\label{sec:Systematics}Systematic uncertainties}
 
This section describes the sources of systematic uncertainty considered for the cross-section and the asymmetry measurements.
The size of these uncertainties as a function of \modetam{} is provided in Figures~\ref{fig:crosssection_uncertainties} and~\ref{fig:asymmetry_uncertainties}.
The data statistical uncertainties are also shown and are small compared with the total systematic uncertainty.
Table~\ref{tab:xssystbin-abs} lists the \Wpmn{} and \Wmmn{} cross-sections and the asymmetry as a function of the absolute pseudorapidity of the muon,
along with the data statistical uncertainties and dominant systematic uncertainties.
Most sources of systematic uncertainty, described below, are treated as being correlated between the positive and negative muon channels unless otherwise noted,
and therefore their relative impact is reduced for the asymmetry measurement.
 
\begin{figure}[ht]
\centering
\includegraphics[width=0.45\textwidth]{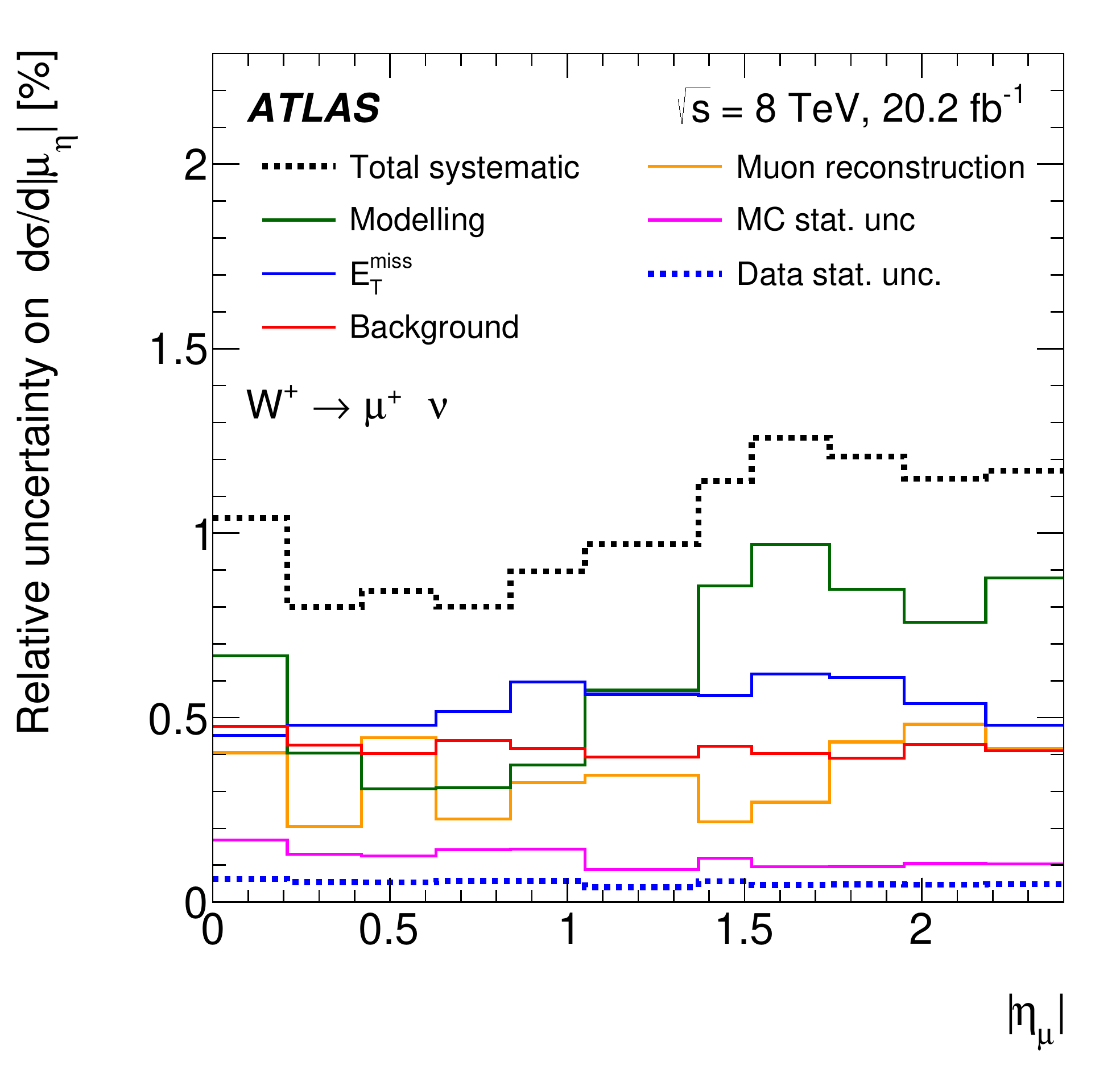}
\includegraphics[width=0.45\textwidth]{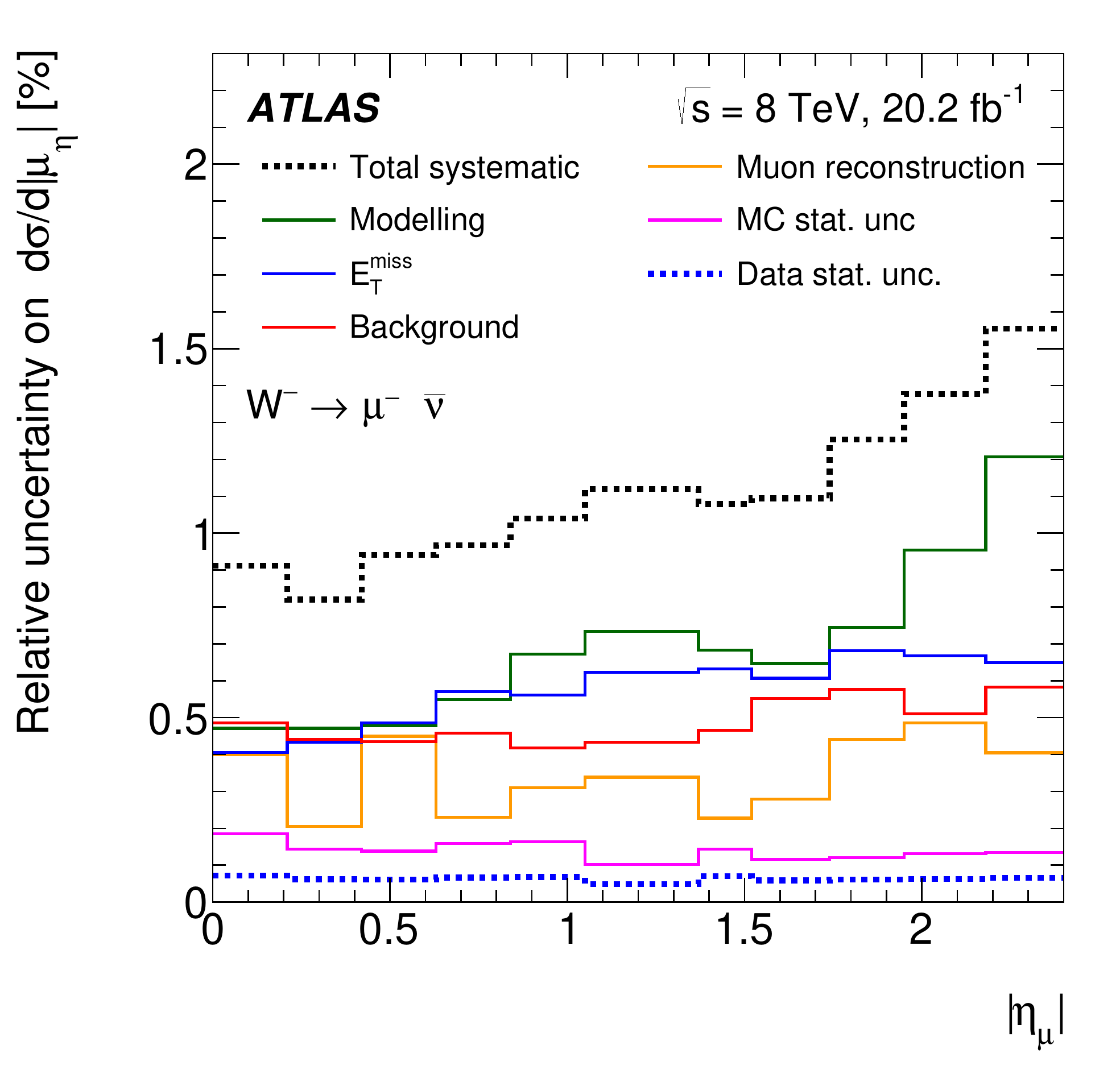}\\
\caption{
The relative systematic uncertainty from each source for the \Wplus{} (left) and \Wminus{} (right) differential cross-sections as a percentage of the differential cross-section.
Also shown are the total systematic and statistical uncertainties.
}
\label{fig:crosssection_uncertainties}
\end{figure}
 
\begin{figure}[ht]
\centering
\includegraphics[width=0.45\textwidth]{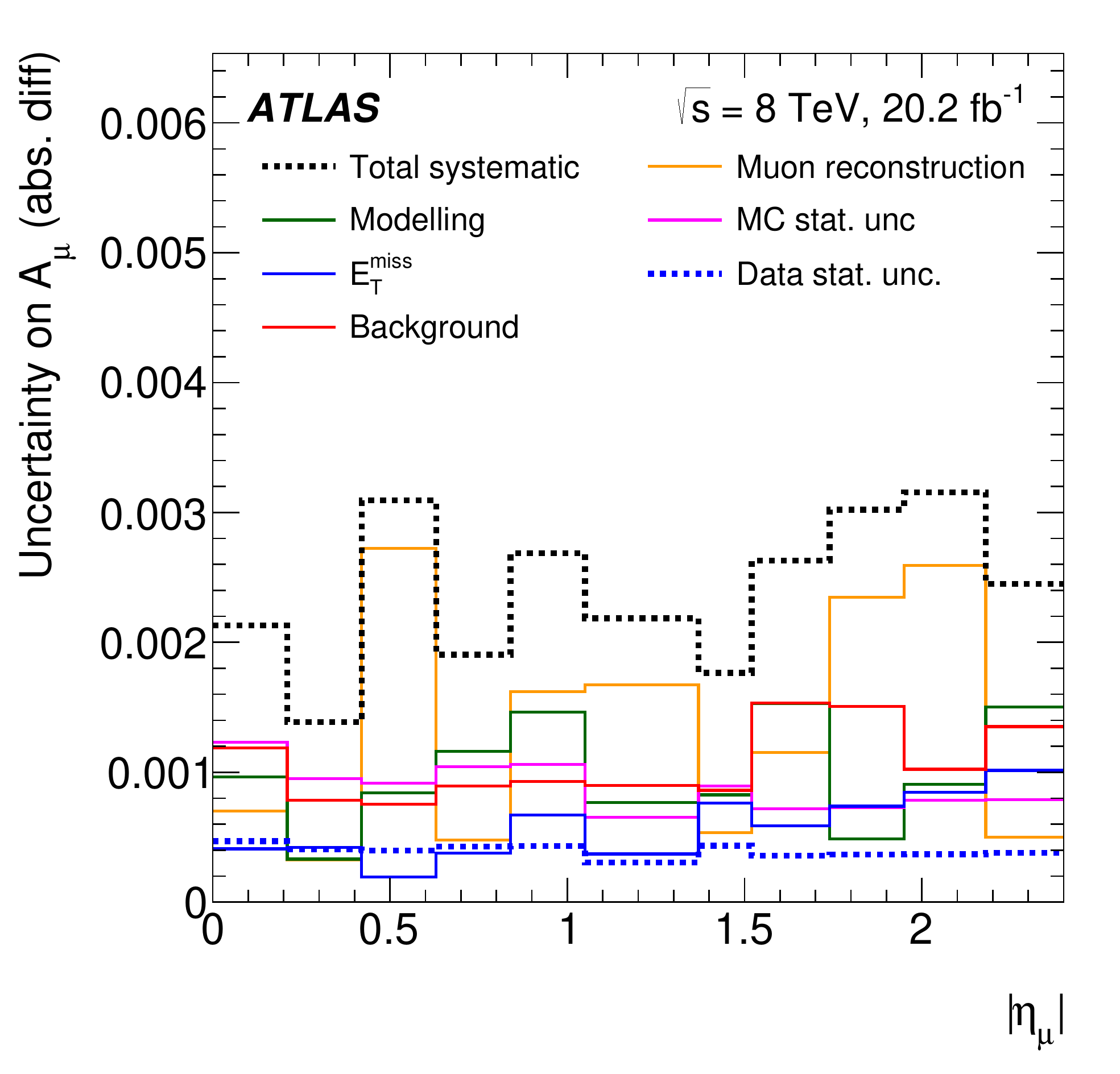}
\caption{
The systematic uncertainty from each source for the \Wboson{} boson charge asymmetry as an absolute difference from the central value.
Also shown are the total systematic and statistical uncertainties.
}
\label{fig:asymmetry_uncertainties}
\end{figure}

Experimental sources of uncertainty are possible mis-modelling of the muon momentum scale, resolution, or charge-dependent sagitta bias
as well as of the reconstruction, trigger, and isolation efficiencies.
Such uncertainties form a small but non-negligible fraction of the total uncertainty, 0.5\% or less of the differential cross-section.
A test is performed to check the compatibility of the cross-sections measured separately in positive and negative muon pseudorapidity; this
is an important cross-check of the correction procedure as the detector is not forward-backward symmetric with respect to the trajectory of a charged particle.
A further small uncertainty (up to 0.4\% depending on the \modetam{} bin) is added to cover the small differences observed.
This uncertainty is treated as being uncorrelated between charges and propagated to the asymmetry measurement.
The above uncertainties are combined in the column labelled `Muon Reconstruction' in Table~\ref{tab:xssystbin-abs}.
Additional sources of uncertainty are the mis-modelling of the pile-up event activity and of the primary vertex longitudinal position, both of which are small.
 
Uncertainties from the mis-modelling of the missing transverse momentum also contribute substantially to the total systematic uncertainty in the cross-section,
although they are reduced for the asymmetry measurements.
These include mis-modelling in the jet energy scale and resolution, as well as of the momentum balance between the soft term and the total transverse momentum of the hard objects in \Zmm{} calibration events~\cite{PERF-2014-04}.
The muon-related uncertainties in the missing transverse momentum are treated as being fully correlated with those of the signal muon
and are part of the muon systematic uncertainties.
The sum of all soft-term uncertainties is the largest or second largest contributor (depending on the \modetam{} bin) to the total uncertainty in the differential cross-sections
but is significantly less important for the asymmetry measurement.
The hard-term uncertainties are small for both the cross-section and asymmetry measurements.
The soft-term and hard-term uncertainties are assumed to be uncorrelated with each other.
 
Uncertainties due to the mis-modelling of the background processes are also considered.
For the backgrounds modelled with MC simulations, these are estimated by varying their
normalisation within theoretical uncertainties and observing the effect on the final measurements.
The cross-section uncertainty for the \Zmm{} process and the \Wtaunu{} process is 5\%~\cite{ATL-PHYS-PUB-2017-006}.
The cross-section uncertainty for \ttbar{} production is 6\%~\cite{ttbarXsec1,ttbarXsec2,ttbarXsec3,ttbarXsec4,ttbarXsec5,ttbarXsec6,ttbarXsec7,},
and for single top production 7\%~\cite{Kidonakis:2010tc,Kidonakis:2011wy,Kidonakis:2010ux}.
The cross-section uncertainty for diboson production is 5\% for $WW$ and $ZZ$ production and 7\% for $WZ$ production~\cite{ATL-PHYS-PUB-2017-005}.
 
As mentioned in Section~\ref{sec:Backgrounds}, there are three components of the uncertainty in the multijet
background normalisation.
The two correlated uncertainties are larger for the cross-section measurements (totalling around 0.5\%),
whilst the effect of the statistical component is larger for the asymmetry measurement.
 
The statistical uncertainty
due to the limited number of MC events for the other backgrounds and for the signal process is small for the cross-section measurement (less than 0.2\%)
but becomes significant for the asymmetry measurement (around one third of the total uncertainty),
as it is completely uncorrelated between positive and negative channels.
 
Theoretical sources of uncertainty arise from the choice of PDF interfaced to the \Powheg{}+\Pythia{}8 signal MC simulations (the \texttt{CT10} PDF set).
This uncertainty is estimated by reweighting the \Powheg{}+\Pythia{}8 events to the nominal values of the \texttt{CT14}~\cite{Dulat:2015mca} and \texttt{MSTW2008nlo68cl}~\cite{Martin:2009iq} PDF sets using the LHAPDF interface~\cite{Whalley:2005nh}, taking the difference to the nominal values and adding in quadrature.
The PDF uncertainty is small for both the fiducial cross-section and asymmetry measurements.
 
The alternative~\Sherpa{} signal sample is used to estimate an uncertainty from \Powheg{}+\Pythia{}8 related to the modelling of the matrix elements that impact kinematics,
as well as the underlying event activity and hadronisation (which affects the \MET{} measurement).
The difference between the dressed-level $C_W$ values obtained using the \Powheg{}+\Pythia{}8 and \Sherpa{} simulations is statistically significant and assigned as a systematic uncertainty.
This is labelled as `Modelling' in Table~\ref{tab:xssystbin-abs}.
The dressed-level is defined by combining the four-momentum of each muon after photon FSR with that of photons radiated within a cone defined by $\Delta R = 0.1$ around the muon.
This is one of the largest systematic uncertainties for both the cross-section (up to 1\% at large \modetam{}) and the asymmetry measurements.
The uncertainty in the luminosity is 1.9\%~\cite{DAPR-2013-01}.
For the asymmetry, the uncertainty in the luminosity only affects the asymmetry measurement through negligible effects in the background estimation, and it is therefore considered fully correlated between the \Wplus{} and \Wminus{} samples.

\begin{table}
\caption{Cross-sections (differential in \mueta{}) and asymmetry, as a function of \modetam{}. The central values are provided along with the statistical and dominant systematic uncertainties: the data statistical uncertainty (Data Stat.), the \MET{} uncertainty, the uncertainties related to muon reconstruction (Muon Reco.), those related to the background, those from MC statistics (MC Stat.), and modelling uncertainties. The uncertainties of the cross-sections are given in percent and those of the asymmetry as an absolute difference from the nominal.}
\label{tab:xssystbin-abs}
\centering
\resizebox{0.95\textwidth}{!}{
\begin{tabular}{ cc c c c c c c }
\toprule
\modetam{}  &   & \makecell{Data \\ Stat.}  &  \MET{}  & \makecell{Muon \\ Reco.}  & \makecell{Back- \\ ground}  & \makecell{MC \\ Stat.}  & Modelling \\
\midrule
& $d\sigma_{W_{\mu^{+}}}/d\eta_{\mu}$ [pb] &  \multicolumn{5}{c}{} \\
\midrule
0.00--0.21 & 630.3 & 0.06 & 0.45 & 0.40 & 0.48 & 0.17 & 0.67\\
0.21--0.42 & 635.2 & 0.05 & 0.48 & 0.21 & 0.42 & 0.13 & 0.40\\
0.42--0.63 & 641.6 & 0.05 & 0.48 & 0.45 & 0.40 & 0.13 & 0.31\\
0.63--0.84 & 638.1 & 0.06 & 0.52 & 0.23 & 0.44 & 0.14 & 0.31\\
0.84--1.05 & 642.8 & 0.06 & 0.60 & 0.32 & 0.42 & 0.14 & 0.37\\
1.05--1.37 & 654.7 & 0.04 & 0.56 & 0.34 & 0.39 & 0.09 & 0.57\\
1.37--1.52 & 661.0 & 0.06 & 0.56 & 0.22 & 0.42 & 0.12 & 0.86\\
1.52--1.74 & 662.3 & 0.05 & 0.62 & 0.27 & 0.40 & 0.10 & 0.97\\
1.74--1.95 & 661.8 & 0.05 & 0.61 & 0.43 & 0.39 & 0.10 & 0.85\\
1.95--2.18 & 657.5 & 0.05 & 0.54 & 0.48 & 0.43 & 0.10 & 0.76\\
2.18--2.40 & 641.5 & 0.05 & 0.48 & 0.42 & 0.41 & 0.10 & 0.88\\
\midrule
& $d\sigma_{W_{\mu^{-}}}/d\eta_{\mu}$ [pb] &  \multicolumn{5}{c}{} \\
\midrule
0.00 - 0.21 & 493.7 & 0.07 & 0.41 & 0.40 & 0.49 & 0.18 & 0.47\\
0.21 - 0.42 & 489.6 & 0.06 & 0.43 & 0.20 & 0.44 & 0.14 & 0.47\\
0.42 - 0.63 & 485.8 & 0.06 & 0.49 & 0.45 & 0.43 & 0.14 & 0.48\\
0.63 - 0.84 & 474.3 & 0.07 & 0.57 & 0.23 & 0.46 & 0.16 & 0.55\\
0.84 - 1.05 & 465.0 & 0.07 & 0.56 & 0.31 & 0.42 & 0.16 & 0.67\\
1.05 - 1.37 & 455.7 & 0.05 & 0.62 & 0.34 & 0.43 & 0.10 & 0.73\\
1.37 - 1.52 & 439.7 & 0.07 & 0.63 & 0.23 & 0.47 & 0.14 & 0.68\\
1.52 - 1.74 & 427.3 & 0.06 & 0.61 & 0.28 & 0.55 & 0.12 & 0.65\\
1.74 - 1.95 & 410.2 & 0.06 & 0.68 & 0.44 & 0.58 & 0.12 & 0.74\\
1.95 - 2.18 & 389.1 & 0.06 & 0.67 & 0.49 & 0.51 & 0.13 & 0.95\\
2.18 - 2.40 & 368.3 & 0.07 & 0.65 & 0.40 & 0.58 & 0.13 & 1.21\\
\midrule
& $A_{\mu}$ &  \multicolumn{5}{c}{} \\
\midrule
0.00 - 0.21 & 0.1215 & 0.0005 & 0.0004 & 0.0007 & 0.0012 & 0.0012 & 0.0010\\
0.21 - 0.42 & 0.1294 & 0.0004 & 0.0004 & 0.0003 & 0.0008 & 0.0010 & 0.0003\\
0.42 - 0.63 & 0.1383 & 0.0004 & 0.0002 & 0.0027 & 0.0008 & 0.0009 & 0.0008\\
0.63 - 0.84 & 0.1473 & 0.0004 & 0.0004 & 0.0005 & 0.0009 & 0.0010 & 0.0012\\
0.84 - 1.05 & 0.1605 & 0.0004 & 0.0007 & 0.0016 & 0.0009 & 0.0011 & 0.0015\\
1.05 - 1.37 & 0.1792 & 0.0003 & 0.0004 & 0.0017 & 0.0009 & 0.0007 & 0.0008\\
1.37 - 1.52 & 0.2011 & 0.0004 & 0.0008 & 0.0005 & 0.0009 & 0.0009 & 0.0008\\
1.52 - 1.74 & 0.2156 & 0.0004 & 0.0006 & 0.0012 & 0.0015 & 0.0007 & 0.0015\\
1.74 - 1.95 & 0.2347 & 0.0004 & 0.0007 & 0.0023 & 0.0015 & 0.0007 & 0.0005\\
1.95 - 2.18 & 0.2565 & 0.0004 & 0.0008 & 0.0026 & 0.0010 & 0.0008 & 0.0009\\
2.18 - 2.40 & 0.2706 & 0.0004 & 0.0010 & 0.0005 & 0.0014 & 0.0008 & 0.0015\\
\bottomrule
\end{tabular}
}
\end{table}
 
 
\section{Theoretical predictions}
The \Wplus{} and \Wminus{} integrated and differential cross-sections and the \Wboson{} boson charge asymmetry
are compared with theoretical predictions from an optimised version of the DYNNLO generator~\cite{Catani:2009sm},
which simulates initial-state QCD corrections to NNLO accuracy at leading order in the electroweak
couplings with parameters set according to the $G_{\mu}$ scheme~\cite{Hollik:1988ii}.
The input parameters (the Fermi constant $G_\mathrm{F}$, the masses and widths of the \Wboson{} and \Zboson{} bosons, and the CKM matrix elements) are taken from Ref.~\cite{Patrignani:2016xqp}.
The DYNNLO predictions are calculated with the PDF sets from \texttt{CT14 NNLO}~\cite{Dulat:2015mca}, \texttt{ATLASepWZ2016}~\cite{STDM-2012-20}, \texttt{HERAPDF2.0}~\cite{Abramowicz:2015mha}, \texttt{NNPDF3.1}~\cite{Ball:2017nwa}, \texttt{PDF4LHC15}~\cite{Butterworth:2015oua} and \texttt{MMHT2014 NNLO}~\cite{Harland-Lang:2014zoa}.
The renormalisation and factorisation scales are set equal to the invariant mass of the muon--neutrino pair.
 
Uncertainties in the DYNNLO prediction due the choice of scales are evaluated by varying the factorisation and renormalisation scales independently by factors of 0.5 and 2 from their nominal values.
The uncertainty on the \texttt{CT14 NNLO} prediction is evaluated using the corresponding PDF error sets.
 
\section{Results}
 
The measured integrated fiducial cross-sections multiplied by the branching fraction for the decay into a muon and a neutrino are listed in Table~\ref{tab:integrated-total-combined}, along with the associated uncertainties which are dominated by the 1.9\% uncertainty in the luminosity.
Also provided is the sum of the \Wpmn{} and \Wmmn{} integrated cross-sections and their ratio,
the total uncertainties on which are 2.1\% and 0.3\% respectively.
The data are compared with the NNLO predictions from DYNNLO, the total uncertainties on which are dominated by the component from the PDF uncertainty. Data and theory agree well and within the uncertainties, although the ratio measurements differ by approximately two standard deviations.
The results presented in Table~\ref{tab:integrated-total-combined} are consistent, within about one standard deviation, with those measured using \Wboson{} boson decays in the electron channel in data at a centre-of-mass energy of 8~\TeV{}~\cite{STDM-2016-14}.
 
\begin{table}
\caption{The measured fiducial production cross-sections times branching ratio for \Wpmn{} and \Wmmn{}, their sum, and their ratio for both data and the predictions from DYNNLO (\texttt{CT14 NNLO} PDF set).}
\label{tab:integrated-total-combined}
\centering
\begin{center}
\begin{tabular}{ ll }
\toprule
\multicolumn{2}{c}{ Data } \\
\midrule
$\sigma$ (\Wpmn{}) [pb] & 3110 $\pm$ 0.5 (stat.) $\pm$ 28 (syst.) $\pm$ 59 (lumi.) \\
$\sigma$ (\Wmmn{}) [pb] & 2137 $\pm$ 0.4 (stat.) $\pm$ 21 (syst.) $\pm$ 41 (lumi.) \\
Sum [pb] & 5247 $\pm$ 0.6 (stat.) $\pm$ 49 (syst.) $\pm$ 100 (lumi.) \\
Ratio & 1.4558 $\pm$ 0.0004 (stat.) $\pm$ 0.0040 (syst.) \\
\midrule
\multicolumn{2}{c}{ DYNNLO (\texttt{CT14 NNLO} PDF set) } \\
\midrule
$\sigma$ (\Wpmn{}) [pb] & 3015 $\pm$ \hphantom{0}92 (PDF) $\pm$ 15 (scale) \\
$\sigma$ (\Wmmn{}) [pb] & 2105 $\pm$ \hphantom{0}53 (PDF) $\pm$ 10 (scale) \\
Sum [pb] & 5120 $\pm$ 140 (PDF) $\pm$ 23 (scale) \\
Ratio & 1.4320 $\pm$ 0.0100 (PDF) $\pm$ 0.0007 (scale) \\
\bottomrule
\end{tabular}
\end{center}
\end{table}
 
The \Wplus{} and \Wminus{} fiducial cross-sections, differential in muon pseudorapidity, multiplied by the branching fraction for the decay into a muon and a neutrino, are shown as a function of absolute muon pseudorapidity in Figure~\ref{fig:differential-crosssection}.
These are presented with systematic and total uncertainties.
The data statistical uncertainties are smaller than the size of the markers.
The cross-section values are detailed in Table~\ref{tab:xssystbin-abs}.
The measured cross-sections are compared with theoretical predictions obtained using DYNNLO.
 
In the top plots of Figure~\ref{fig:differential-crosssection}, DYNNLO is shown with its associated total theoretical uncertainty,
and from the PDF uncertainty (evaluated with the \texttt{CT14 NNLO} error sets).
The component from the PDF uncertainty dominates the total and is shown separately.
The data precision is similar to the intrinsic theoretical uncertainty from scale variations but is much higher than the uncertainty from the PDF.
Therefore the data are useful for constraining and evaluating the performance of different PDF sets.
In the bottom plots the data are compared with the central values of each PDF set described above.
The statistical uncertainties of the DYNNLO predictions are indicated by error bars.
 
\begin{figure}[ht]
\centering
\includegraphics[width=0.45\textwidth]{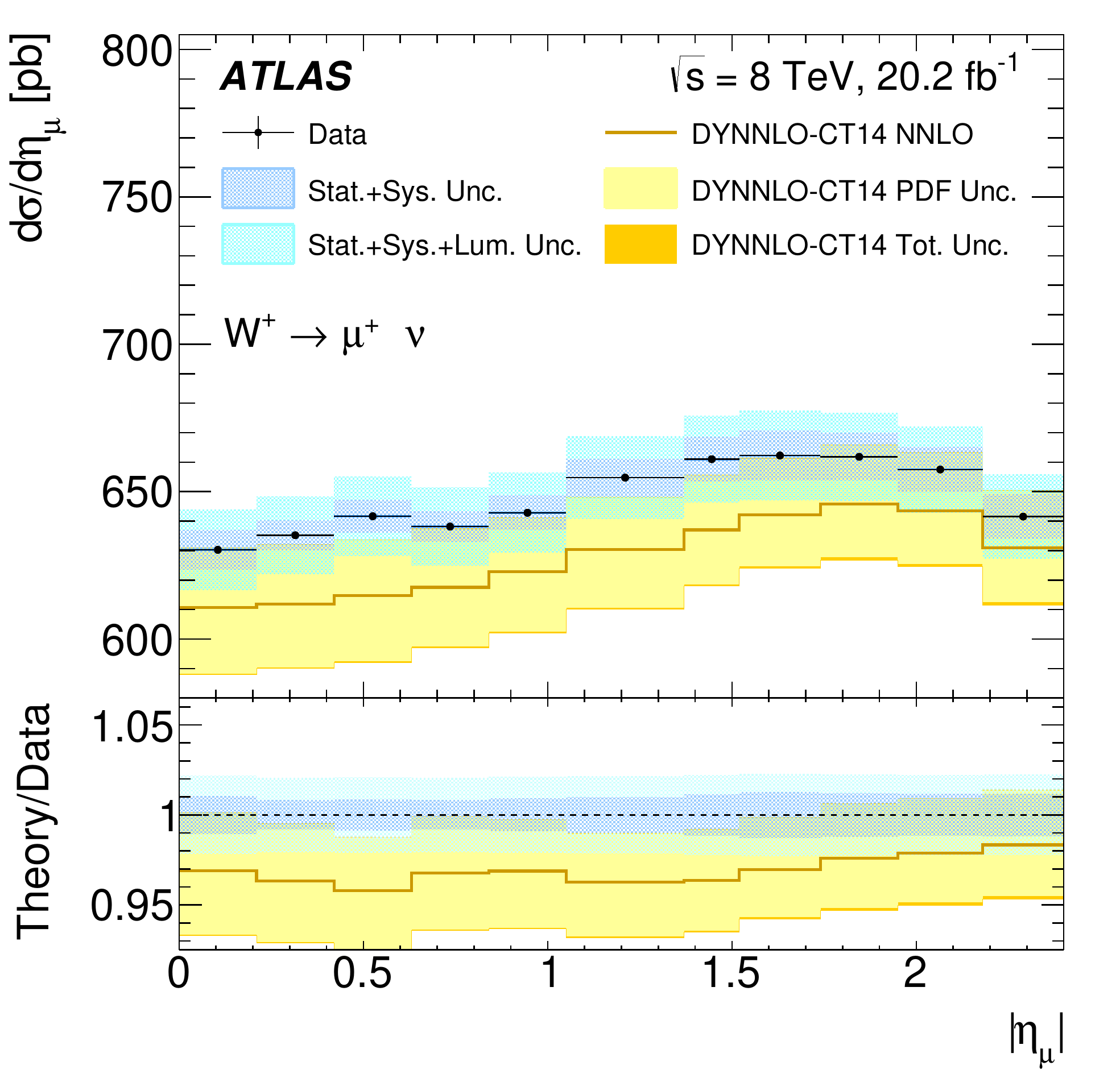}
\includegraphics[width=0.45\textwidth]{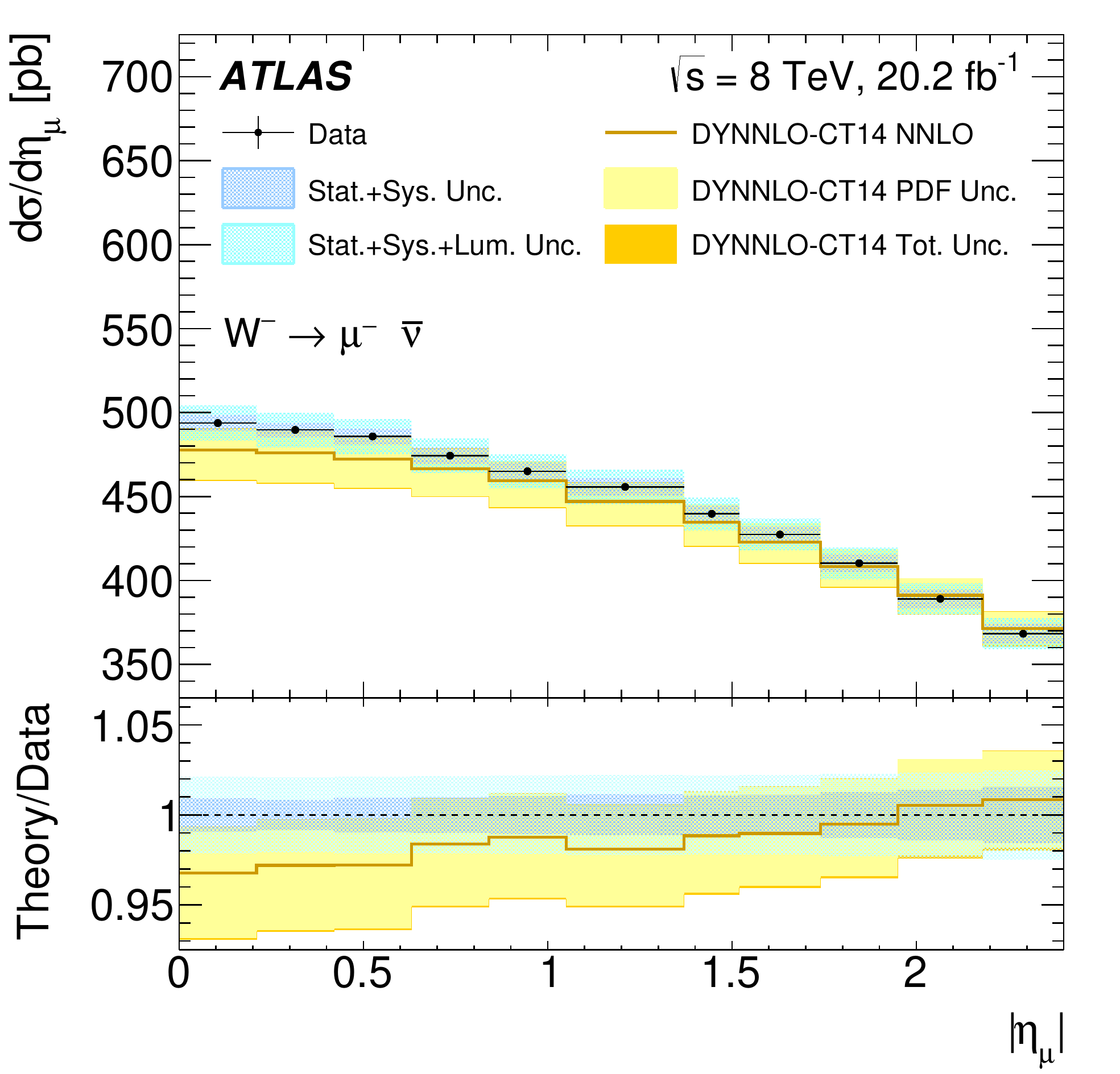}
 
\includegraphics[width=0.45\textwidth]{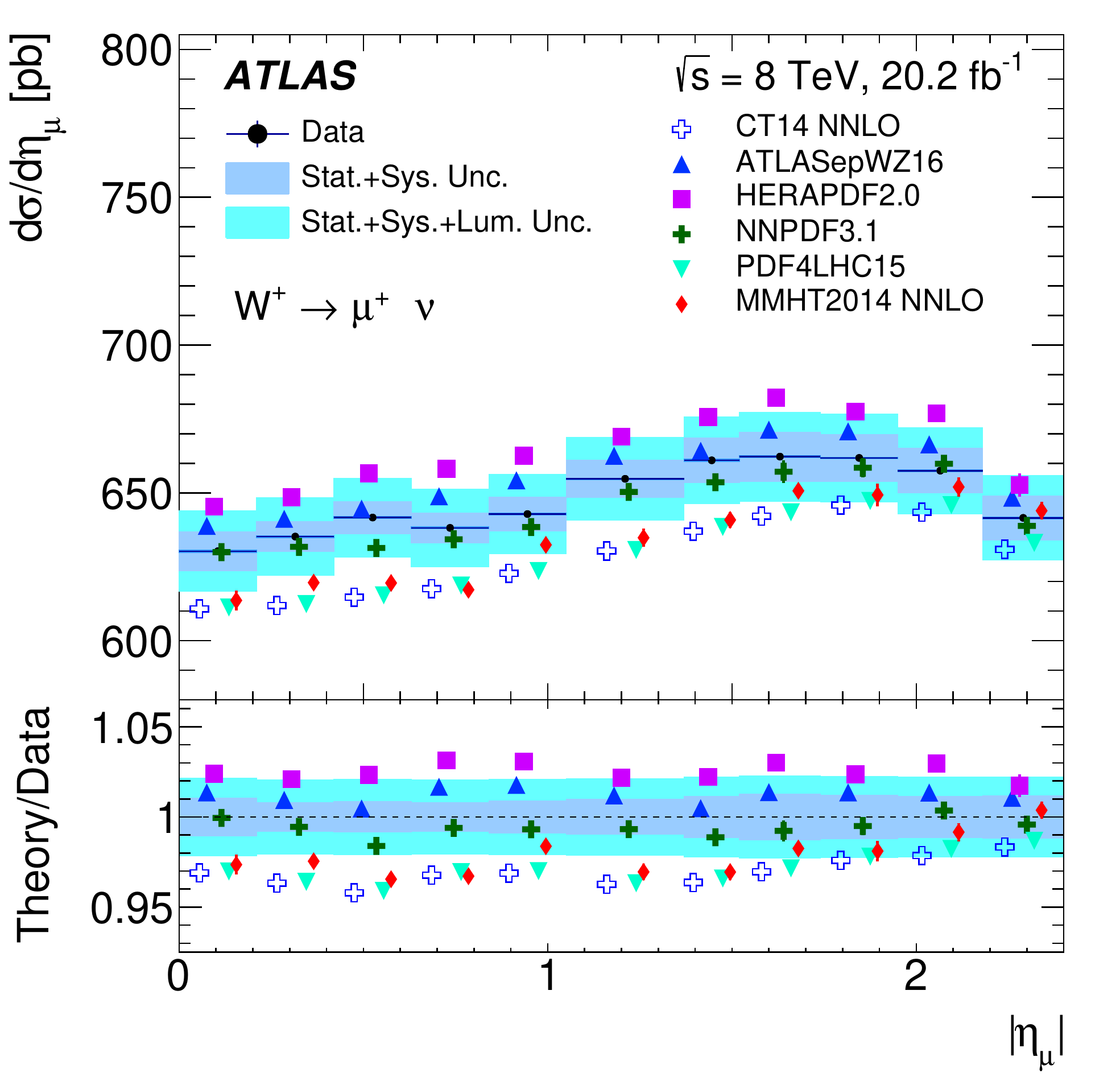}
\includegraphics[width=0.45\textwidth]{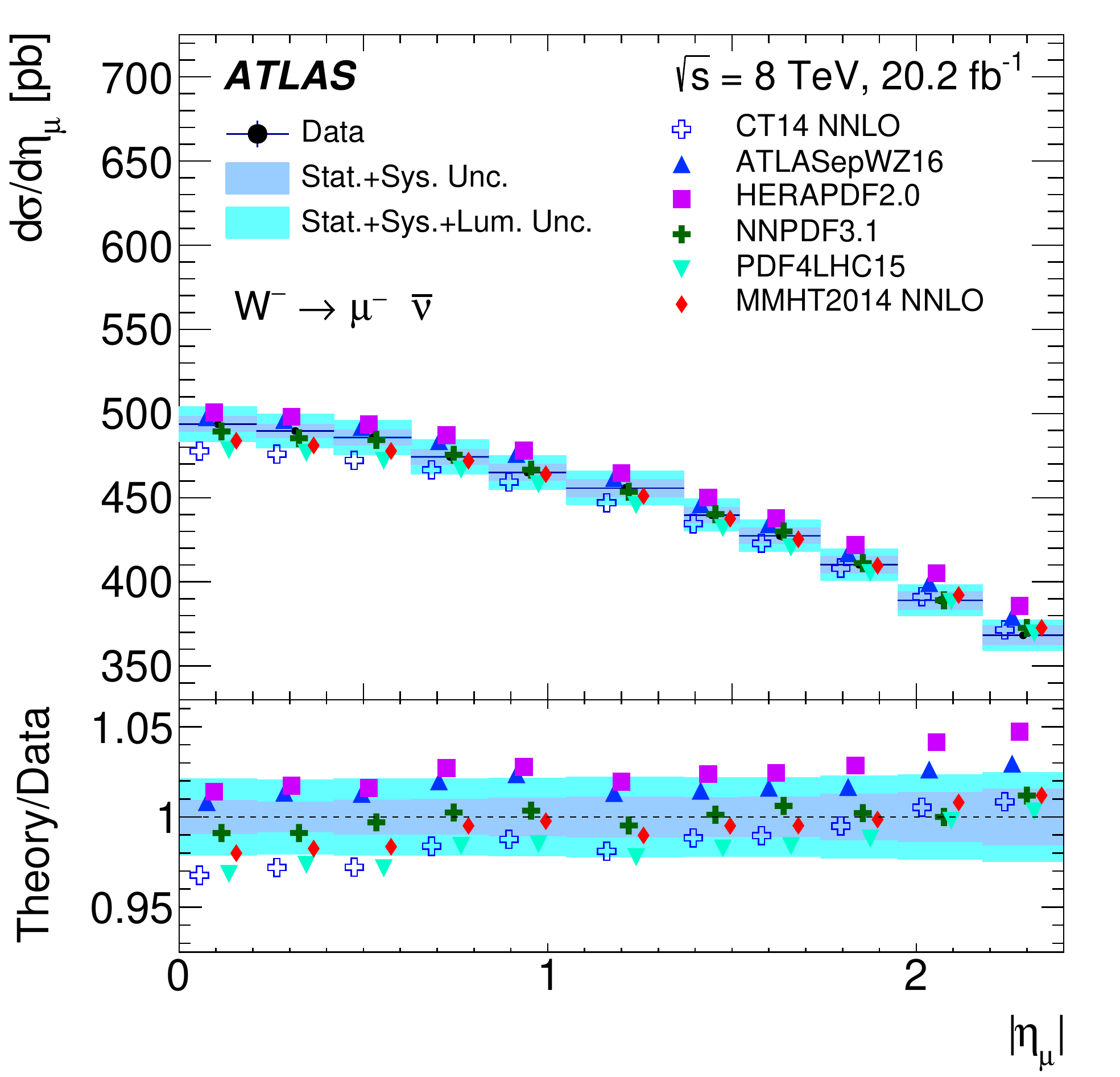}
\caption{
The $\Wplus$ (left) and $\Wminus$ (right) fiducial cross-sections, differential in muon pseudorapidity multiplied by the branching fraction for the decay into a muon and a neutrino are shown as a function of the absolute muon pseudorapidity. The data are presented with systematic and total uncertainties (the data statistical uncertainties are smaller than the size of the markers) and are compared with the predictions from DYNNLO. In the top two plots, the \texttt{CT14 NNLO} PDF set is used, and DYNNLO is shown with its associated total theoretical uncertainty. In the bottom two plots, the data are compared with the central values of six different PDF sets described in the text. The statistical uncertainties of the DYNNLO predictions are indicated by error bars. The ratios of the data to the corresponding prediction are shown in the lower panels.
}
\label{fig:differential-crosssection}
\end{figure}
 
\begin{figure}[ht]
\centering
\includegraphics[width=0.45\textwidth]{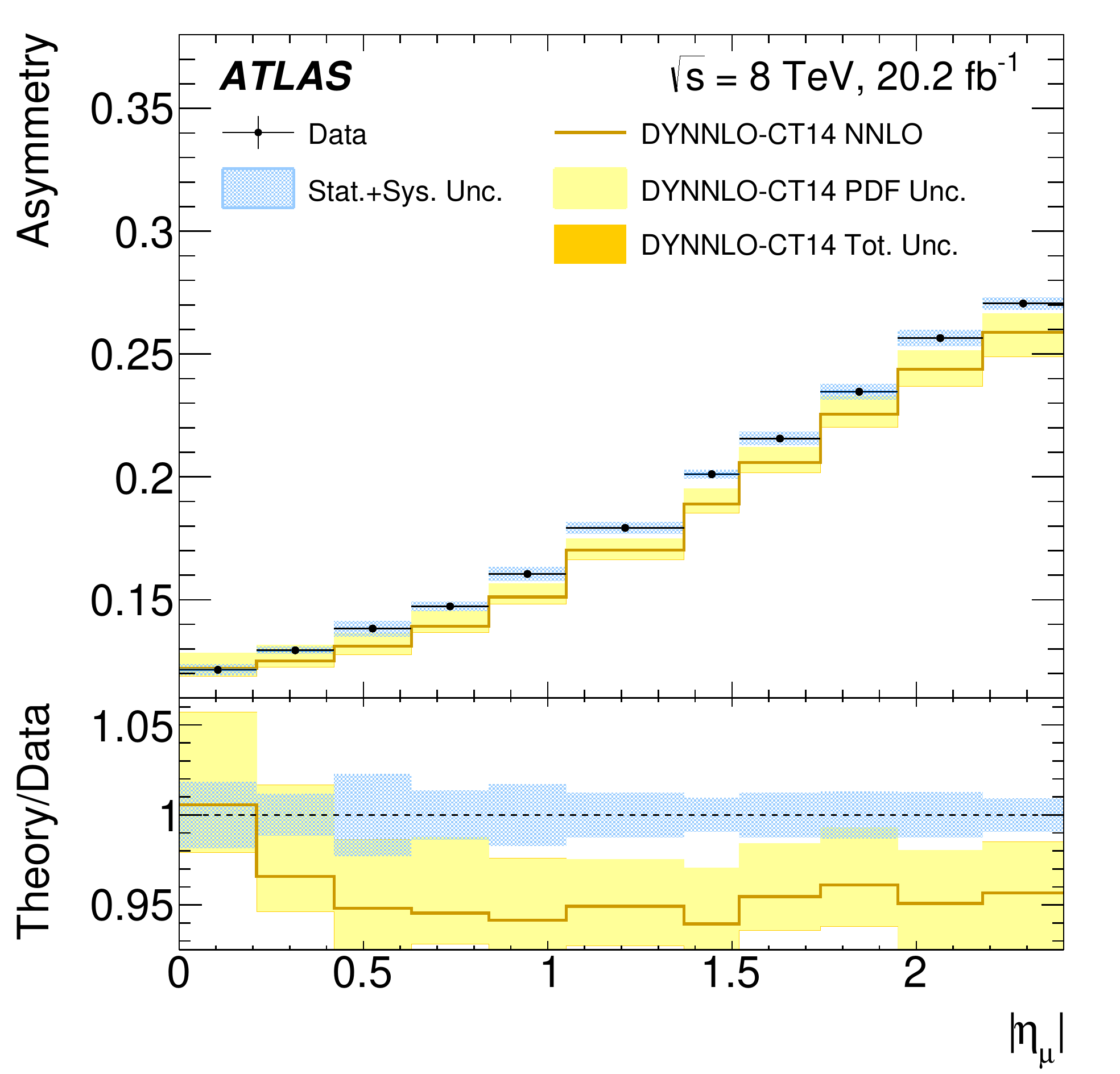}
\includegraphics[width=0.45\textwidth]{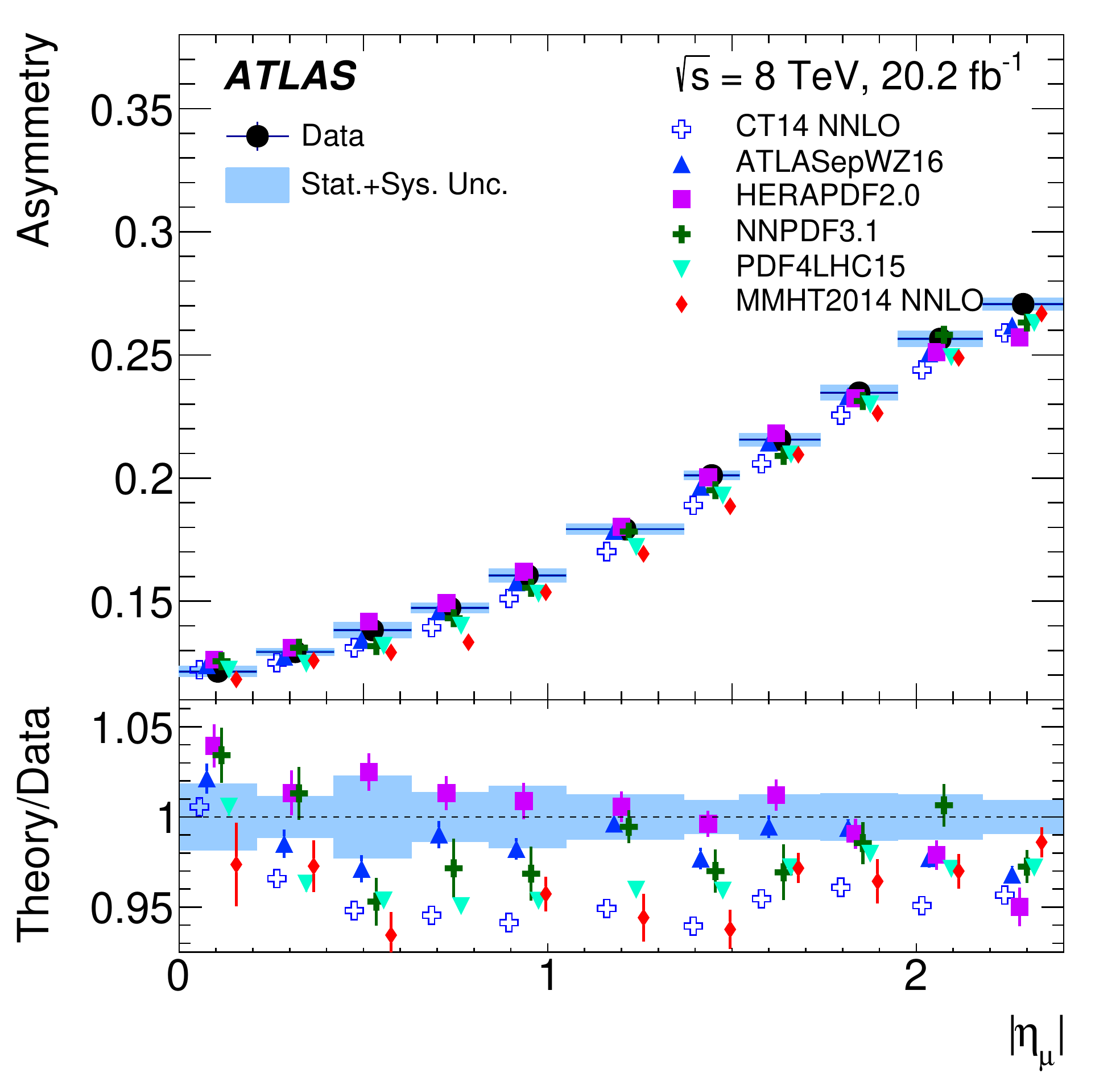}
\caption{
The \Wboson{} boson charge asymmetry as a function of absolute muon pseudorapidity. The data are presented with systematic and total uncertainties (the data statistical uncertainties are smaller than the size of the markers). In the left plot, the data are compared with the prediction from DYNNLO in which the \texttt{CT14 NNLO} PDF set is used. The DYNNLO prediction is also shown with its associated total theoretical uncertainty, along with the component from the PDF set. In the right plot, the data are compared with the central prediction from DYNNLO produced using a selection of PDFs. The statistical uncertainties of the DYNNLO predictions are indicated by error bars. The ratios of the data to the corresponding prediction are shown in the lower panels.
}
\label{fig:asymmetry}
\end{figure}
 
The measured \Wboson{} boson charge asymmetry as a function of absolute muon pseudorapidity is presented in Figure~\ref{fig:asymmetry}.
The values are detailed in Table~\ref{tab:xssystbin-abs}.
Again the data are shown with its total systematic uncertainty.
In Figure~\ref{fig:asymmetry} (left) the data are compared with the prediction from DYNNLO in which the \texttt{CT14 NNLO} PDF set is used.
The DYNNLO prediction is also shown with its associated total theoretical uncertainty, along with the component from the PDF set, which dominates.
In Figure~\ref{fig:asymmetry} (right) the data are compared with the central prediction from the six PDF sets considered.
The statistical uncertainties of the DYNNLO predictions are indicated by error bars.
The ratios of the data to the corresponding prediction are shown in the lower panels.
The comparison with \texttt{ATLASepWZ2016} and \texttt{NNPDF3.1} is of particular interest as both include information from the ATLAS 7~\TeV{} measurement~\cite{STDM-2012-20}, which is expected to be largely uncorrelated with the current data being presented.
It is observed that its central value is generally closer to the data than the alternatives, other than \texttt{HERAPDF2.0} which performs about as well.

\FloatBarrier
 
\section{Conclusion}
 
Fiducial cross-sections for \Wpmn{} and \Wmmn{} and the \Wboson{} boson charge asymmetry
are measured differentially as a function of the absolute muon pseudorapidity
using $20.2~\ifb$ of data from proton--proton collisions at a centre-of-mass energy of 8~\TeV{} with the ATLAS experiment at the LHC.
The muon and neutrino transverse momenta are required to be greater than 25~\GeV{} and the \Wboson{} boson transverse mass to be greater than 40~\GeV{}.
A precision of 0.8--1.5\% is achieved for the cross-section values, depending on the pseudorapidity, whilst an uncertainty between 0.002 and 0.003 (in absolute units) is obtained for the asymmetry.
The integrated fiducial $W^{\pm}$ production cross-sections are also determined.
The measurements are compared to predictions at NNLO accuracy in QCD computed with the DYNNLO program.
The precision of the measurement is better than both the uncertainties on the PDF sets as well as the spread between different sets,
showing the sensitivity of the measurement to discriminate between them and serve as input to improve the knowledge on the proton structure.

 
\section*{Acknowledgements}
 
 
We thank CERN for the very successful operation of the LHC, as well as the
support staff from our institutions without whom ATLAS could not be
operated efficiently.
 
We acknowledge the support of ANPCyT, Argentina; YerPhI, Armenia; ARC, Australia; BMWFW and FWF, Austria; ANAS, Azerbaijan; SSTC, Belarus; CNPq and FAPESP, Brazil; NSERC, NRC and CFI, Canada; CERN; CONICYT, Chile; CAS, MOST and NSFC, China; COLCIENCIAS, Colombia; MSMT CR, MPO CR and VSC CR, Czech Republic; DNRF and DNSRC, Denmark; IN2P3-CNRS, CEA-DRF/IRFU, France; SRNSFG, Georgia; BMBF, HGF, and MPG, Germany; GSRT, Greece; RGC, Hong Kong SAR, China; ISF and Benoziyo Center, Israel; INFN, Italy; MEXT and JSPS, Japan; CNRST, Morocco; NWO, Netherlands; RCN, Norway; MNiSW and NCN, Poland; FCT, Portugal; MNE/IFA, Romania; MES of Russia and NRC KI, Russian Federation; JINR; MESTD, Serbia; MSSR, Slovakia; ARRS and MIZ\v{S}, Slovenia; DST/NRF, South Africa; MINECO, Spain; SRC and Wallenberg Foundation, Sweden; SERI, SNSF and Cantons of Bern and Geneva, Switzerland; MOST, Taiwan; TAEK, Turkey; STFC, United Kingdom; DOE and NSF, United States of America. In addition, individual groups and members have received support from BCKDF, CANARIE, CRC and Compute Canada, Canada; COST, ERC, ERDF, Horizon 2020, and Marie Sk{\l}odowska-Curie Actions, European Union; Investissements d' Avenir Labex and Idex, ANR, France; DFG and AvH Foundation, Germany; Herakleitos, Thales and Aristeia programmes co-financed by EU-ESF and the Greek NSRF, Greece; BSF-NSF and GIF, Israel; CERCA Programme Generalitat de Catalunya, Spain; The Royal Society and Leverhulme Trust, United Kingdom.
 
The crucial computing support from all WLCG partners is acknowledged gratefully, in particular from CERN, the ATLAS Tier-1 facilities at TRIUMF (Canada), NDGF (Denmark, Norway, Sweden), CC-IN2P3 (France), KIT/GridKA (Germany), INFN-CNAF (Italy), NL-T1 (Netherlands), PIC (Spain), ASGC (Taiwan), RAL (UK) and BNL (USA), the Tier-2 facilities worldwide and large non-WLCG resource providers. Major contributors of computing resources are listed in Ref.~\cite{ATL-GEN-PUB-2016-002}.
 
 
\printbibliography

@techreport{ATL-PHYS-PUB-2014-021,
      title         = "{ATLAS Run 1 Pythia8 tunes}",
      institution   = "CERN",
      address       = "Geneva",
      number        = "ATL-PHYS-PUB-2014-021",
      month         = "Nov",
      year          = "2014",
      reportNumber  = "ATL-PHYS-PUB-2014-021",
      url           = "https://cds.cern.ch/record/1966419",
}

@article{Aaltonen:2009ta,
  author =	 "{CDF Collaboration}",
  title =	 "{Direct Measurement of the $W$ Production Charge
                  Asymmetry in $p\bar{p}$ Collisions at $\sqrt{s} =
                  1.96$ TeV}",
  collaboration ="CDF",
  journal =	 "Phys. Rev. Lett.",
  volume =	 "102",
  year =	 "2009",
  pages =	 "181801",
  doi =		 "10.1103/PhysRevLett.102.181801",
  eprint =	 "0901.2169",
  archivePrefix ="arXiv",
  primaryClass = "hep-ex",
  reportNumber = "FERMILAB-PUB-09-017-E",
  SLACcitation = "%%CITATION = ARXIV:0901.2169;%%"
}

@article{Abazov:2013rja2,
  author =	 "{D0 Collaboration}",
  title =	 "{Measurement of the muon charge asymmetry in
                  $p\bar{p}$ $\to$ W+X $\to$ $\mu$$\nu$ + X events at
                  \sqs=1.96~\TeV}",
  collaboration ="D0",
  journal =	 "Phys. Rev. D",
  volume =	 "88",
  year =	 "2013",
  pages =	 "091102",
  doi =		 "10.1103/PhysRevD.88.091102",
  eprint =	 "1309.2591",
  archivePrefix ="arXiv",
  primaryClass = "hep-ex",
  reportNumber = "FERMILAB-PUB-13-361-E",
  SLACcitation = "%%CITATION = ARXIV:1309.2591;%%"
}

@article{D0:2014kma,
  author =	 "{D0 Collaboration}",
  title =	 "{Measurement of the electron charge asymmetry in
                  $p\bar{p}\rightarrow W+X \rightarrow
                  e\nu +X$ decays in $p\bar{p}$
                  collisions at $\sqrt{s}=1.96$ TeV}",
  collaboration ="D0",
  journal =	 "Phys. Rev. D",
  volume =	 "91",
  year =	 "2015",
  number =	 "3",
  pages =	 "032007",
  doi =		 "10.1103/PhysRevD.91.032007",
  related =      "D0:2014kma_erratum",
  relatedstring = "Erratum:",
  eprint =	 "1412.2862",
  archivePrefix ="arXiv",
  primaryClass = "hep-ex",
  reportNumber = "FERMILAB-PUB-14-514-E",
  SLACcitation = "%%CITATION = ARXIV:1412.2862;%%"
}

@article{Sjostrand:2006za,
  author =	 "Sj{\"o}strand, Torbjorn and Mrenna, Stephen and Skands,
                  Peter Z.",
  title =	 "{PYTHIA 6.4 physics and manual}",
  journal =	 "JHEP",
  volume =	 "05",
  pages =	 "026",
  doi =		 "10.1088/1126-6708/2006/05/026",
  year =	 "2006",
  eprint =	 "hep-ph/0603175",
  archivePrefix ="arXiv",
  reportNumber = "FERMILAB-PUB-06-052-CD-T, LU-TP-06-13",
  SLACcitation = "%%CITATION = HEP-PH/0603175;%%",
}

@article{Corcella:2000bw,
  author =	 "Corcella, G. and others",
  title =	 "{HERWIG 6: An event generator for hadron emission
                  reactions with interfering gluons (including
                  supersymmetric processes)}",
  journal =	 "JHEP",
  volume =	 "01",
  pages =	 "010",
  doi =		 "10.1088/1126-6708/2001/01/010",
  year =	 "2001",
  eprint =	 "hep-ph/0011363",
  archivePrefix ="arXiv",
  reportNumber = "CAVENDISH-HEP-99-03, CERN-TH-2000-284,
                  RAL-TR-2000-048",
  SLACcitation = "%%CITATION = HEP-PH/0011363;%%",
}

@article{Whalley:2005nh,
  author =	 "Whalley, M.R. and Bourilkov, D. and Group, R.C.",
  title =	 "{The Les Houches Accord PDFs (LHAPDF) and LHAGLUE}",
  year =	 "2005",
  eprint =	 "hep-ph/0508110",
  archivePrefix ="arXiv",
  SLACcitation = "%%CITATION = HEP-PH/0508110;%%",
}

@article{Catani:2009sm,
  author =	 "Catani, Stefano and Cieri, Leandro and Ferrera,
                  Giancarlo and de Florian, Daniel and Grazzini,
                  Massimiliano",
  title =	 "{Vector Boson Production at Hadron Colliders: 
                  A Fully Exclusive QCD Calculation at Next-to-Next-to-Leading Order}",
  journal =	 "Phys. Rev. Lett.",
  volume =	 "103",
  pages =	 "082001",
  doi =		 "10.1103/PhysRevLett.103.082001",
  year =	 "2009",
  eprint =	 "0903.2120",
  archivePrefix ="arXiv",
  primaryClass = "hep-ph",
  SLACcitation = "%%CITATION = ARXIV:0903.2120;%%",
}

@article{Catani:2007vq,
      author         = "Catani, Stefano and Grazzini, Massimiliano",
      title          = "{Next-to-Next-to-Leading-Order Subtraction Formalism in Hadron Collisions and
                        its Application to Higgs-Boson Production at the LHC}",
      journal        = "Phys. Rev. Lett.",
      volume         = "98",
      year           = "2007",
      pages          = "222002",
      doi            = "10.1103/PhysRevLett.98.222002",
      eprint         = "hep-ph/0703012",
      archivePrefix  = "arXiv",
      SLACcitation   = "%%CITATION = HEP-PH/0703012;%%"
}

@article{Alioli:2008gx,
  author =	 "Alioli, Simone and Nason, Paolo and Oleari, Carlo
                  and Re, Emanuele",
  title =	 "{NLO vector-boson production matched with shower in
                  POWHEG}",
  journal =	 "JHEP",
  volume =	 "07",
  pages =	 "060",
  doi =		 "10.1088/1126-6708/2008/07/060",
  year =	 "2008",
  eprint =	 "0805.4802",
  archivePrefix ="arXiv",
  primaryClass = "hep-ph",
  reportNumber = "NSF-KITP-08-74",
  SLACcitation = "%%CITATION = ARXIV:0805.4802;%%",
}

@article{Gleisberg:2008ta,
  author =	 "Gleisberg, T. and others",
  title =	 "{Event generation with SHERPA 1.1}",
  journal =	 "JHEP",
  volume =	 "02",
  pages =	 "007",
  doi =		 "10.1088/1126-6708/2009/02/007",
  year =	 "2009",
  eprint =	 "0811.4622",
  archivePrefix ="arXiv",
  primaryClass = "hep-ph",
  reportNumber = "FERMILAB-PUB-08-477-T, SLAC-PUB-13420, ZU-TH-17-08,
                  DCPT-08-138, IPPP-08-69, EDINBURGH-2008-30,
                  MCNET-08-14",
  SLACcitation = "%%CITATION = ARXIV:0811.4622;%%",
}

@article{Sjostrand:2007gs,
  author =	 "Sj{\"o}strand, Torbjorn and Mrenna, Stephen and Skands,
                  Peter Z.",
  title =	 "{A brief introduction to PYTHIA 8.1}",
  journal =	 "Comput. Phys. Commun.",
  volume =	 "178",
  pages =	 "852-867",
  doi =		 "10.1016/j.cpc.2008.01.036",
  year =	 "2008",
  eprint =	 "0710.3820",
  archivePrefix ="arXiv",
  primaryClass = "hep-ph",
  reportNumber = "CERN-LCGAPP-2007-04, LU-TP-07-28,
                  FERMILAB-PUB-07-512-CD-T",
  SLACcitation = "%%CITATION = ARXIV:0710.3820;%%",
}

@article{Golonka:2005pn,
  author =	 "Golonka, Piotr and Was, Zbigniew",
  title =	 "{PHOTOS Monte Carlo: a precision tool for QED
                  corrections in $Z$ and $W$ decays}",
  journal =	 "Eur. Phys. J. C",
  volume =	 "45",
  year =	 "2006",
  pages =	 "97-107",
  doi =		 "10.1140/epjc/s2005-02396-4",
  eprint =	 "hep-ph/0506026",
  archivePrefix ="arXiv",
  reportNumber = "IFJPAN-V-05-01, CERN-PH-TH-2005-091",
  SLACcitation = "%%CITATION = HEP-PH/0506026;%%"
}

@article{Mangano:2002ea,
  author =	 "Mangano, Michelangelo L. and Moretti, Mauro and
                  Piccinini, Fulvio and Pittau, Roberto and Polosa,
                  Antonio D.",
  title =	 "{ALPGEN, a generator for hard multiparton processes
                  in hadronic collisions}",
  journal =	 "JHEP",
  volume =	 "07",
  pages =	 "001",
  doi =		 "10.1088/1126-6708/2003/07/001",
  year =	 "2003",
  eprint =	 "hep-ph/0206293",
  archivePrefix ="arXiv",
  reportNumber = "CERN-TH-2002-129, FTN-T-2002-06",
  SLACcitation = "%%CITATION = HEP-PH/0206293;%%",
}

@article{Butterworth:1996zw,
  author =	 "Butterworth, J.M. and Forshaw, Jeffrey R. and
                  Seymour, M.H.",
  title =	 "{Multiparton Interactions in Photoproduction at HERA}",
  journal =	 "Z. Phys. C",
  volume =	 "72",
  pages =	 "637-646",
  doi =		 "10.1007/s002880050286",
  year =	 "1996",
  eprint =	 "hep-ph/9601371",
  archivePrefix ="arXiv",
  reportNumber = "CERN-TH-96-82, M-C-TH-96-05, UCL-HEP-96-02",
  SLACcitation = "%%CITATION = HEP-PH/9601371;%%",
}

@article{Frixione:2002ik,
  author =	 "Frixione, Stefano and Webber, Bryan R.",
  title =	 "{Matching NLO QCD computations and parton shower
                  simulations}",
  journal =	 "JHEP",
  volume =	 "06",
  pages =	 "029",
  doi =		 "10.1088/1126-6708/2002/06/029",
  year =	 "2002",
  eprint =	 "hep-ph/0204244",
  archivePrefix ="arXiv",
  reportNumber = "CAVENDISH-HEP-02-01, LAPTH-905-02, GEF-TH-2-2002",
  SLACcitation = "%%CITATION = HEP-PH/0204244;%%",
}

@article{Kersevan:2004yg,
  author =	 "Kersevan, Borut Paul and Richter-Was, Elzbieta",
  title =	 "{The Monte Carlo event generator AcerMC versions 2.0
                  to 3.8 with interfaces to PYTHIA 6.4, HERWIG 6.5 and
                  ARIADNE 4.1}",
  journal =	 "Comput. Phys. Commun.",
  volume =	 "184",
  pages =	 "919-985",
  doi =		 "10.1016/j.cpc.2012.10.032",
  year =	 "2013",
  eprint =	 "hep-ph/0405247",
  archivePrefix ="arXiv",
  reportNumber = "TPJU-6-2004",
  SLACcitation = "%%CITATION = HEP-PH/0405247;%%",
}

@article{Agostinelli:2002hh,
  author =	 "Agostinelli, S. and others",
  title =	 "{GEANT4 --- a simulation toolkit}",
  collaboration ="GEANT4",
  journal =	 "Nucl. Instrum. Meth. Phys. Res. A",
  volume =	 "506",
  pages =	 "250-303",
  doi =		 "10.1016/S0168-9002(03)01368-8",
  year =	 "2003",
  reportNumber = "SLAC-PUB-9350, FERMILAB-PUB-03-339",
  SLACcitation = "%%CITATION = NUIMA,A506,250;%%",
}

@article{Pumplin:2002vw,
  author =	 "Pumplin, J. and Stump, D. R. and Huston, J. and Lai,
                  H.  L. and Nadolsky, Pavel M. and Tung, W. K.",
  title =	 "{New Generation of Parton Distributions with
                  Uncertainties from Global QCD Analysis}",
  journal =	 "JHEP",
  volume =	 "07",
  year =	 "2002",
  pages =	 "012",
  doi =		 "10.1088/1126-6708/2002/07/012",
  eprint =	 "hep-ph/0201195",
  archivePrefix ="arXiv",
  reportNumber = "MSU-HEP-011101",
  SLACcitation = "%%CITATION = HEP-PH/0201195;%%"
}

@article{Aad:2010ah,
  author =	 "{ATLAS Collaboration}",
  title =	 "{The ATLAS Simulation Infrastructure}",
  collaboration ="ATLAS",
  journal =	 "Eur. Phys. J. C",
  volume =	 "70",
  year =	 "2010",
  pages =	 "823-874",
  doi =		 "10.1140/epjc/s10052-010-1429-9",
  eprint =	 "1005.4568",
  archivePrefix ="arXiv",
  primaryClass = "physics.ins-det",
  SLACcitation = "%%CITATION = ARXIV:1005.4568;%%"
}

@article{Frixione:2007nw,
      author         = "Frixione, Stefano and Nason, Paolo and Ridolfi, Giovanni",
      title          = "{A Positive-weight next-to-leading-order Monte Carlo for
                        heavy flavour hadroproduction}",
      journal        = "JHEP",
      volume         = "09",
      year           = "2007",
      pages          = "126",
      doi            = "10.1088/1126-6708/2007/09/126",
      eprint         = "0707.3088",
      archivePrefix  = "arXiv",
      primaryClass   = "hep-ph",
      reportNumber   = "BICOCCA-FT-07-12, GEF-TH-19-2007",
      SLACcitation   = "%%CITATION = ARXIV:0707.3088;%%"
}

@article{Alioli:2010xd,
  author =	 "Alioli, Simone and Nason, Paolo and Oleari, Carlo
                  and Re, Emanuele",
  title =	 "{A general framework for implementing NLO
                  calculations in shower Monte Carlo programs: the
                  POWHEG BOX}",
  journal =	 "JHEP",
  volume =	 "06",
  year =	 "2010",
  pages =	 "043",
  doi =		 "10.1007/JHEP06(2010)043",
  eprint =	 "1002.2581",
  archivePrefix ="arXiv",
  primaryClass = "hep-ph",
  reportNumber = "DESY-10-018, SFB-CPP-10-22, IPPP-10-11, DCPT-10-22",
  SLACcitation = "%%CITATION = ARXIV:1002.2581;%%"
}

@article{Dulat:2015mca,
      author         = "Dulat, Sayipjamal and Hou, Tie-Jiun and Gao, Jun and
                        Guzzi, Marco and Huston, Joey and Nadolsky, Pavel and
                        Pumplin, Jon and Schmidt, Carl and Stump, Daniel and Yuan,
                        C. P.",
      title          = "{New parton distribution functions from a global analysis
                        of quantum chromodynamics}",
      journal        = "Phys. Rev. D",
      volume         = "93",
      year           = "2016",
      number         = "3",
      pages          = "033006",
      doi            = "10.1103/PhysRevD.93.033006",
      eprint         = "1506.07443",
      archivePrefix  = "arXiv",
      primaryClass   = "hep-ph",
      SLACcitation   = "%%CITATION = ARXIV:1506.07443;%%"
}

@article{Kidonakis:2010tc,
      author         = "Kidonakis, Nikolaos",
      title          = "{Next-to-next-to-leading logarithm resummation for s-channel single top quark
                        production}",
      journal        = "Phys. Rev. D",
      volume         = "81",
      year           = "2010",
      pages          = "054028",
      doi            = "10.1103/PhysRevD.81.054028",
      eprint         = "1001.5034",
      archivePrefix  = "arXiv",
      primaryClass   = "hep-ph",
      SLACcitation   = "%%CITATION = ARXIV:1001.5034;%%"
}

@article{Kidonakis:2011wy,
      author         = "Kidonakis, Nikolaos",
      title          = "{Next-to-next-to-leading-order collinear and soft gluon
                        corrections for t-channel single top quark production}",
      journal        = "Phys. Rev. D",
      volume         = "83",
      year           = "2011",
      pages          = "091503",
      doi            = "10.1103/PhysRevD.83.091503",
      eprint         = "1103.2792",
      archivePrefix  = "arXiv",
      primaryClass   = "hep-ph",
      SLACcitation   = "%%CITATION = ARXIV:1103.2792;%%"
}

@article{Kidonakis:2010ux,
      author         = "Kidonakis, Nikolaos",
      title          = "{Two-loop soft anomalous dimensions for single top quark
                        associated production with a $W^{-}$ or $H^{-}$}",
      journal        = "Phys. Rev. D",
      volume         = "82",
      year           = "2010",
      pages          = "054018",
      doi            = "10.1103/PhysRevD.82.054018",
      eprint         = "1005.4451",
      archivePrefix  = "arXiv",
      primaryClass   = "hep-ph",
      SLACcitation   = "%%CITATION = ARXIV:1005.4451;%%"
}

@article{Martin:2009iq,
      author         = "Martin, A. D. and Stirling, W. J. and Thorne, R. S. and
                        Watt, G.",
      title          = "{Parton distributions for the LHC}",
      journal        = "Eur. Phys. J. C",
      volume         = "63",
      year           = "2009",
      pages          = "189-285",
      doi            = "10.1140/epjc/s10052-009-1072-5",
      eprint         = "0901.0002",
      archivePrefix  = "arXiv",
      primaryClass   = "hep-ph",
      reportNumber   = "IPPP-08-95, DCPT-08-190, CAVENDISH-HEP-08-16",
      SLACcitation   = "%%CITATION = ARXIV:0901.0002;%%"
}

@article{Abramowicz:2015mha,
      author         = "{H1 and ZEUS Collaborations}",
      title          = "{Combination of measurements of inclusive deep inelastic
                        ${e^{\pm }p}$ scattering cross sections and QCD analysis
                        of HERA data}",
      collaboration  = "H1, ZEUS",
      journal        = "Eur. Phys. J. C",
      volume         = "75",
      year           = "2015",
      number         = "12",
      pages          = "580",
      doi            = "10.1140/epjc/s10052-015-3710-4",
      eprint         = "1506.06042",
      archivePrefix  = "arXiv",
      primaryClass   = "hep-ex",
      reportNumber   = "DESY-15-039",
      SLACcitation   = "%%CITATION = ARXIV:1506.06042;%%"
}

@article{Ball:2017nwa,
      author         = "Ball, Richard D. and others",
      title          = "{Parton distributions from high-precision collider data}",
      collaboration  = "NNPDF",
      journal        = "Eur. Phys. J. C",
      volume         = "77",
      year           = "2017",
      number         = "10",
      pages          = "663",
      doi            = "10.1140/epjc/s10052-017-5199-5",
      eprint         = "1706.00428",
      archivePrefix  = "arXiv",
      primaryClass   = "hep-ph",
      reportNumber   = "CAVENDISH-HEP-17-06, CERN-TH-2017-077, EDINBURGH-2017-08,
                        NIKHEF-2017-006, OUTP-17-04P, TIF-UNIMI-2017-3",
      SLACcitation   = "%%CITATION = ARXIV:1706.00428;%%"
}

@article{Butterworth:2015oua,
      author         = "Butterworth, Jon and others",
      title          = "{PDF4LHC recommendations for LHC Run II}",
      journal        = "J. Phys. G",
      volume         = "43",
      year           = "2016",
      pages          = "023001",
      doi            = "10.1088/0954-3899/43/2/023001",
      eprint         = "1510.03865",
      archivePrefix  = "arXiv",
      primaryClass   = "hep-ph",
      reportNumber   = "OUTP-15-17P, SMU-HEP-15-12, TIF-UNIMI-2015-14,
                        LCTS-2015-27, CERN-PH-TH-2015-249",
      SLACcitation   = "%%CITATION = ARXIV:1510.03865;%%"
}

@article{Harland-Lang:2014zoa,
      author         = "Harland-Lang, L. A. and Martin, A. D. and Motylinski, P.
                        and Thorne, R. S.",
      title          = "{Parton distributions in the LHC era: MMHT 2014 PDFs}",
      journal        = "Eur. Phys. J. C",
      volume         = "75",
      year           = "2015",
      number         = "5",
      pages          = "204",
      doi            = "10.1140/epjc/s10052-015-3397-6",
      eprint         = "1412.3989",
      archivePrefix  = "arXiv",
      primaryClass   = "hep-ph",
      reportNumber   = "LCTS-2014-47, IPPP-14-97, DCPT-14-194",
      SLACcitation   = "%%CITATION = ARXIV:1412.3989;%%"
}

@article{Aaij:2014wba,
      author         = "{LHCb Collaboration}",
      title          = "{Measurement of the forward $W$ boson cross-section in
                        $pp$ collisions at $\sqrt{s} = 7 {\textrm \, TeV}$}",
      collaboration  = "LHCb",
      journal        = "JHEP",
      volume         = "12",
      year           = "2014",
      pages          = "079",
      doi            = "10.1007/JHEP12(2014)079",
      eprint         = "1408.4354",
      archivePrefix  = "arXiv",
      primaryClass   = "hep-ex",
      reportNumber   = "LHCB-PAPER-2014-033, CERN-PH-EP-2014-175",
      SLACcitation   = "%%CITATION = ARXIV:1408.4354;%%"
}

@article{Aaij:2015zlq,
      author         = "{LHCb Collaboration}",
      title          = "{Measurement of forward W and Z boson production in $pp$
                        collisions at $ \sqrt{s}=8 $ TeV}",
      collaboration  = "LHCb",
      journal        = "JHEP",
      volume         = "01",
      year           = "2016",
      pages          = "155",
      doi            = "10.1007/JHEP01(2016)155",
      eprint         = "1511.08039",
      archivePrefix  = "arXiv",
      primaryClass   = "hep-ex",
      reportNumber   = "LHCB-PAPER-2015-049, CERN-PH-EP-2015-301",
      SLACcitation   = "%%CITATION = ARXIV:1511.08039;%%"
}

@article{Aaij:2016qqz,
      author         = "{LHCb Collaboration}",
      title          = "{Measurement of forward $W\to e\nu$ production in $pp$
                        collisions at $\sqrt{s}=8\,$TeV}",
      collaboration  = "LHCb",
      journal        = "JHEP",
      volume         = "10",
      year           = "2016",
      pages          = "030",
      doi            = "10.1007/JHEP10(2016)030",
      eprint         = "1608.01484",
      archivePrefix  = "arXiv",
      primaryClass   = "hep-ex",
      reportNumber   = "LHCB-PAPER-2016-024, CERN-EP-2016-179",
      SLACcitation   = "%%CITATION = ARXIV:1608.01484;%%"
}

@article{Todesco:2017nnk,
      author         = "Todesco, E. and Wenninger, J.",
      title          = "{Large Hadron Collider momentum calibration and
                        accuracy}",
      journal        = "Phys. Rev. Accel. Beams",
      volume         = "20",
      year           = "2017",
      number         = "8",
      pages          = "081003",
      doi            = "10.1103/PhysRevAccelBeams.20.081003",
      SLACcitation   = "%%CITATION = INSPIRE-1618103;%%"
}

@article{Hollik:1988ii,
      author         = "Hollik, W. F. L.",
      title          = "{Radiative Corrections in the Standard Model and Their
                        Role for Precision Tests of the Electroweak Theory}",
      journal        = "Fortsch. Phys.",
      volume         = "38",
      year           = "1990",
      pages          = "165-260",
      doi            = "10.1002/prop.2190380302",
      reportNumber   = "DESY-88-188",
      SLACcitation   = "%%CITATION = FPYKA,38,165;%%"
}

@article{Patrignani:2016xqp,
      author         = "Patrignani, C. and others",
      title          = "{Review of Particle Physics}",
      collaboration  = "Particle Data Group",
      journal        = "Chin. Phys.",
      volume         = "C40",
      year           = "2016",
      number         = "10",
      pages          = "100001",
      doi            = "10.1088/1674-1137/40/10/100001",
      SLACcitation   = "%%CITATION = CHPHD,C40,100001;%%"
}

@article{ttbarXsec1,
     author         = "Cacciari, Matteo and Czakon, Michal and Mangano,
                       Michelangelo and Mitov, Alexander and Nason, Paolo",
     title          = "{Top-pair production at hadron colliders with
                       next-to-next-to-leading logarithmic soft-gluon
                       resummation}",
     journal        = "Phys. Lett. B",
     volume         = "710",
     year           = "2012",
     pages          = "612-622",
     doi            = "10.1016/j.physletb.2012.03.013",
     eprint         = "1111.5869",
     archivePrefix  = "arXiv",
     primaryClass   = "hep-ph",
     reportNumber   = "CERN-PH-TH-2011-277, TTK-11-54",
     SLACcitation   = "%%CITATION = ARXIV:1111.5869;%%"
}

@article{ttbarXsec2,
     author         = "Beneke, M. and Falgari, P. and Klein, S. and Schwinn, C.",
     title          = "{Hadronic top-quark pair production with NNLL threshold
                       resummation}",
     journal        = "Nucl. Phys. B",
     volume         = "855",
     year           = "2012",
     pages          = "695-741",
     doi            = "10.1016/j.nuclphysb.2011.10.021",
     eprint         = "1109.1536",
     archivePrefix  = "arXiv",
     primaryClass   = "hep-ph",
     reportNumber   = "TTK-11-38, ITP-UU-11-26, SPIN-11-19, FR-PHENO-2011-015,
                       SFB-CPP-11-49",
     SLACcitation   = "%%CITATION = ARXIV:1109.1536;%%"
}

@article{ttbarXsec3,
  author =	 {B\"arnreuther, Peter and Czakon, Micha\l{} and
                  Mitov, Alexander},
  title =	 "{Percent-Level-Precision Physics at the Tevatron:
                  First Genuine Next-to-Next-to-Leading Order QCD Corrections to $q \bar{q} \to
                  t \bar{t} + X$}",
  journal =	 "Phys. Rev. Lett.",
  volume =	 "109",
  year =	 "2012",
  pages =	 "132001",
  doi =		 "10.1103/PhysRevLett.109.132001",
  eprint =	 "1204.5201",
  archivePrefix ="arXiv",
  primaryClass = "hep-ph",
  reportNumber = "CERN-PH-TH-2012-092",
  SLACcitation = "%%CITATION = ARXIV:1204.5201;%%"
}

@article{ttbarXsec4,
  author =	 "Czakon, Michal and Mitov, Alexander",
  title =	 "{NNLO corrections to top-pair production at hadron
                  colliders: the all-fermionic scattering channels}",
  journal =	 "JHEP",
  volume =	 "12",
  year =	 "2012",
  pages =	 "054",
  doi =		 "10.1007/JHEP12(2012)054",
  eprint =	 "1207.0236",
  archivePrefix ="arXiv",
  primaryClass = "hep-ph",
  reportNumber = "CERN-PH-TH-2012-181",
  SLACcitation = "%%CITATION = ARXIV:1207.0236;%%"
}

@article{ttbarXsec5,
  author =	 "Czakon, Michal and Mitov, Alexander",
  title =	 "{NNLO corrections to top pair production at hadron
                  colliders: the quark-gluon reaction}",
  journal =	 "JHEP",
  volume =	 "01",
  year =	 "2013",
  pages =	 "080",
  doi =		 "10.1007/JHEP01(2013)080",
  eprint =	 "1210.6832",
  archivePrefix ="arXiv",
  primaryClass = "hep-ph",
  reportNumber = "CERN-PH-TH-2012-286",
  SLACcitation = "%%CITATION = ARXIV:1210.6832;%%"
}

@article{ttbarXsec6,
  author =	 "Czakon, Micha? and Fiedler, Paul and Mitov,
                  Alexander",
  title =	 "{Total Top-Quark Pair-Production Cross Section at
                  Hadron Colliders Through $O(\alpha^4_S)$}",
  journal =	 "Phys. Rev. Lett.",
  volume =	 "110",
  year =	 "2013",
  pages =	 "252004",
  doi =		 "10.1103/PhysRevLett.110.252004",
  eprint =	 "1303.6254",
  archivePrefix ="arXiv",
  primaryClass = "hep-ph",
  reportNumber = "CERN-PH-TH-2013-056, TTK-13-08",
  SLACcitation = "%%CITATION = ARXIV:1303.6254;%%"
}

@article{ttbarXsec7,
  author =	 "Czakon, Michal and Mitov, Alexander",
  title =	 "{Top++: A program for the calculation of the
                  top-pair cross-section at hadron colliders}",
  journal =	 "Comput. Phys. Commun.",
  volume =	 "185",
  year =	 "2014",
  pages =	 "2930",
  doi =		 "10.1016/j.cpc.2014.06.021",
  eprint =	 "1112.5675",
  archivePrefix ="arXiv",
  primaryClass = "hep-ph",
  reportNumber = "CERN-PH-TH-2011-303, TTK-11-58",
  SLACcitation = "%%CITATION = ARXIV:1112.5675;%%"
}

@article{Lai:2010vv,
      author         = "Lai, Hung-Liang and Guzzi, Marco and Huston, Joey and Li,
                        Zhao and Nadolsky, Pavel M. and Pumplin, Jon and Yuan, C.
                        -P.",
      title          = "{New parton distributions for collider physics}",
      journal        = "Phys. Rev. D",
      volume         = "82",
      year           = "2010",
      pages          = "074024",
      doi            = "10.1103/PhysRevD.82.074024",
      eprint         = "1007.2241",
      archivePrefix  = "arXiv",
      primaryClass   = "hep-ph",
      reportNumber   = "MSUHEP-100707, SMU-HEP-10-10",
      SLACcitation   = "%%CITATION = ARXIV:1007.2241;%%"
}

@article{Alekhin:2017kpj,
      author         = "Alekhin, S. and Blümlein, J. and Moch, S. and Pla\v cakyt\. e,
                        R.",
      title          = "{Parton distribution functions, $\alpha_s$, and
                        heavy-quark masses for LHC Run II}",
      journal        = "Phys. Rev. D",
      volume         = "96",
      year           = "2017",
      number         = "1",
      pages          = "014011",
      doi            = "10.1103/PhysRevD.96.014011",
      eprint         = "1701.05838",
      archivePrefix  = "arXiv",
      primaryClass   = "hep-ph",
      reportNumber   = "DESY-16-179, DO-TH-16-13",
      SLACcitation   = "%%CITATION = ARXIV:1701.05838;%%"
}

@article{Aad:2014sca,
      author         = "{ATLAS Collaboration}",
      title          = "{Performance of the ATLAS muon trigger in pp collisions
                        at $\sqrt{s}=8$ TeV}",
      journal        = "Eur. Phys. J. C",
      volume         = "75",
      year           = "2015",
      pages          = "120",
      doi            = "10.1140/epjc/s10052-015-3325-9",
      eprint         = "1408.3179",
      archivePrefix  = "arXiv",
      primaryClass   = "hep-ex",
      reportNumber   = "CERN-PH-EP-2014-154",
      SLACcitation   = "%%CITATION = ARXIV:1408.3179;%%"
}

@Article{PERF-2007-01,
    author         = "{ATLAS Collaboration}",
    title          = "{The ATLAS Experiment at the CERN Large Hadron Collider}",
    journal        = "JINST",
    volume         = "3",
    year           = "2008",
    pages          = "S08003",
    doi            = "10.1088/1748-0221/3/08/S08003",
    primaryClass   = "hep-ex",
}

@Article{SOFT-2010-01,
    author         = "{ATLAS Collaboration}",
    title          = "{The ATLAS Simulation Infrastructure}",
    journal        = "Eur. Phys. J. C",
    volume         = "70",
    year           = "2010",
    pages          = "823",
    doi            = "10.1140/epjc/s10052-010-1429-9",
    eprint         = "1005.4568",
    archivePrefix  = "arXiv",
    primaryClass   = "physics.ins-det",
}

@Article{PERF-2011-02,
    author         = "{ATLAS Collaboration}",
    title          = "{Performance of the ATLAS Trigger System in 2010}",
    journal        = "Eur. Phys. J. C",
    volume         = "72",
    year           = "2012",
    pages          = "1849",
    doi            = "10.1140/epjc/s10052-011-1849-1",
    reportNumber   = "CERN-PH-EP-2011-078",
    eprint         = "1110.1530",
    archivePrefix  = "arXiv",
    primaryClass   = "hep-ex",
}

@Article{STDM-2012-20,
    author         = "{ATLAS Collaboration}",
    title          = "{Precision measurement and interpretation of inclusive \(W^+\), \(W^-\) and \(Z/\gamma^*\) production cross sections with the ATLAS detector}",
    journal        = "Eur. Phys. J. C",
    volume         = "77",
    year           = "2017",
    pages          = "367",
    doi            = "10.1140/epjc/s10052-017-4911-9",
    reportNumber   = "CERN-EP-2016-272",
    eprint         = "1612.03016",
    archivePrefix  = "arXiv",
    primaryClass   = "hep-ex",
}

@Article{DAPR-2013-01,
    author         = "{ATLAS Collaboration}",
    title          = "{Luminosity determination in \(pp\) collisions at \(\sqrt{s} = 8\,\text{TeV}\) using the ATLAS detector at the LHC}",
    journal        = "Eur. Phys. J. C",
    volume         = "76",
    year           = "2016",
    pages          = "653",
    doi            = "10.1140/epjc/s10052-016-4466-1",
    eprint         = "1608.03953",
    archivePrefix  = "arXiv",
    primaryClass   = "hep-ex",
}

@Article{PERF-2014-04,
    author         = "{ATLAS Collaboration}",
    title          = "{Performance of algorithms that reconstruct missing transverse momentum in \(\sqrt{s} = 8\,\text{TeV}\) proton--proton collisions in the ATLAS detector}",
    journal        = "Eur. Phys. J. C",
    volume         = "77",
    year           = "2017",
    pages          = "241",
    doi            = "10.1140/epjc/s10052-017-4780-2",
    reportNumber   = "CERN-EP-2016-134",
    eprint         = "1609.09324",
    archivePrefix  = "arXiv",
    primaryClass   = "hep-ex",
}

@Article{PERF-2014-05,
    author         = "{ATLAS Collaboration}",
    title          = "{Measurement of the muon reconstruction performance of the ATLAS detector using 2011 and 2012 LHC proton--proton collision data}",
    journal        = "Eur. Phys. J. C",
    volume         = "74",
    year           = "2014",
    pages          = "3130",
    doi            = "10.1140/epjc/s10052-014-3130-x",
    reportNumber   = "CERN-PH-EP-2014-151",
    eprint         = "1407.3935",
    archivePrefix  = "arXiv",
    primaryClass   = "hep-ex",
}

@Article{STDM-2014-18,
    author         = "{ATLAS Collaboration}",
    title          = "{Measurement of the \(W\)-boson mass in \(pp\) collisions at \(\sqrt{s} = 7\,\text{TeV}\) with the ATLAS detector}",
    journal        = "Eur. Phys. J. C",
    volume         = "78",
    year           = "2018",
    pages          = "110",
    doi            = "10.1140/epjc/s10052-017-5475-4",
    reportNumber   = "CERN-EP-2016-305",
    eprint         = "1701.07240",
    archivePrefix  = "arXiv",
    primaryClass   = "hep-ex",
}

@Article{STDM-2016-14,
    author         = "{ATLAS Collaboration}",
    title          = "{Measurement of differential cross sections and \(W^{+}/W^{-}\) cross-section ratios for \(W\) boson production in association with jets at \(\sqrt{s} = 8\,\text{TeV}\) with the ATLAS detector}",
    journal        = "JHEP",
    volume         = "05",
    year           = "2018",
    pages          = "077",
    doi            = "10.1007/JHEP05(2018)077",
    reportNumber   = "CERN-EP-2017-213",
    eprint         = "1711.03296",
    archivePrefix  = "arXiv",
    primaryClass   = "hep-ex",
}

@Article{HION-2018-02,
    author         = "{ATLAS Collaboration}",
    title          = "{Measurements of \(W\) and \(Z\) boson production in \(pp\) collisions at \(\sqrt{s} = 5.02\,\text{TeV}\) with the ATLAS detector}",
    journal        = "Eur. Phys. J. C",
    volume         = "79",
    year           = "2019",
    pages          = "128",
    doi            = "10.1140/epjc/s10052-019-6622-x",
    reportNumber   = "CERN-EP-2018-259",
    eprint         = "1810.08424",
    archivePrefix  = "arXiv",
    primaryClass   = "hep-ex",
}

@Booklet{ATL-GEN-PUB-2016-002,
    author         = "{ATLAS Collaboration}",
    title          = "{ATLAS Computing Acknowledgements}",
    howpublished   = "{ATL-GEN-PUB-2016-002}",
    url            = "https://cds.cern.ch/record/2202407",
}

@Article{CMS-SMP-12-001,
    author         = "{CMS Collaboration}",
    title          = "{Measurement of the electron charge asymmetry in inclusive \(W\) production in \(pp\) collisions at \(\sqrt{s} = 7\,\text{TeV}\)}",
    journal        = "Phys. Rev. Lett.",
    volume         = "109",
    year           = "2012",
    pages          = "111806",
    doi            = "10.1103/PhysRevLett.109.111806",
    reportNumber   = "CERN-PH-EP-2012-151",
    eprint         = "1206.2598",
    archivePrefix  = "arXiv",
    primaryClass   = "hep-ex",
}

@Article{CMS-SMP-12-021,
    author         = "{CMS Collaboration}",
    title          = "{Measurement of the muon charge asymmetry in inclusive \(pp \rightarrow W+X\) production at \(\sqrt{s} = 7\,\text{TeV}\) and an improved determination of light parton distribution functions}",
    journal        = "Phys. Rev. D",
    volume         = "90",
    year           = "2014",
    pages          = "032004",
    doi            = "10.1103/PhysRevD.90.032004",
    reportNumber   = "CERN-PH-EP-2013-232",
    eprint         = "1312.6283",
    archivePrefix  = "arXiv",
    primaryClass   = "hep-ex",
}

@Article{CMS-SMP-14-022,
    author         = "{CMS Collaboration}",
    title          = "{Measurement of the differential cross section and charge asymmetry for inclusive \(pp \rightarrow W^\pm + X\) production at \(\sqrt{s} = 8\,\text{TeV}\)}",
    journal        = "Eur. Phys. J. C",
    volume         = "76",
    year           = "2016",
    pages          = "469",
    doi            = "10.1140/epjc/s10052-016-4293-4",
    reportNumber   = "CERN-EP-2016-043",
    eprint         = "1603.01803",
    archivePrefix  = "arXiv",
    primaryClass   = "hep-ex",
}

@Booklet{ATLAS-CONF-2010-069,
    author         = "{ATLAS Collaboration}",
    title          = "{Performance of primary vertex reconstruction in proton--proton collisions at \(\sqrt{s} = 7~\text{TeV}\) in the ATLAS experiment}",
    howpublished   = "{ATLAS-CONF-2010-069}",
    url            = "https://cds.cern.ch/record/1281344",
    year           = "2010",
}

@Booklet{ATL-PHYS-PUB-2012-003,
    author         = "{ATLAS Collaboration}",
    title          = "{Summary of ATLAS Pythia 8 tunes}",
    howpublished   = "{ATL-PHYS-PUB-2012-003}",
    url            = "https://cds.cern.ch/record/1474107",
    year           = "2012",
}

@Booklet{ATL-PHYS-PUB-2017-005,
    author         = "{ATLAS Collaboration}",
    title          = "{Multi-Boson Simulation for \(13~\text{TeV}\) ATLAS Analyses}",
    howpublished   = "{ATL-PHYS-PUB-2017-005}",
    url            = "https://cds.cern.ch/record/2261933",
    year           = "2017",
}

@Booklet{ATL-PHYS-PUB-2017-006,
    author         = "{ATLAS Collaboration}",
    title          = "{ATLAS simulation of boson plus jets processes in Run~2}",
    howpublished   = "{ATL-PHYS-PUB-2017-006}",
    url            = "https://cds.cern.ch/record/2261937",
    year           = "2017",
}

\clearpage 
 
\begin{flushleft}
{\Large The ATLAS Collaboration}

\bigskip

G.~Aad$^\textrm{\scriptsize 101}$,    
B.~Abbott$^\textrm{\scriptsize 128}$,    
D.C.~Abbott$^\textrm{\scriptsize 102}$,    
O.~Abdinov$^\textrm{\scriptsize 13,*}$,    
A.~Abed~Abud$^\textrm{\scriptsize 70a,70b}$,    
K.~Abeling$^\textrm{\scriptsize 53}$,    
D.K.~Abhayasinghe$^\textrm{\scriptsize 93}$,    
S.H.~Abidi$^\textrm{\scriptsize 167}$,    
O.S.~AbouZeid$^\textrm{\scriptsize 40}$,    
N.L.~Abraham$^\textrm{\scriptsize 156}$
% ...
\bigskip
\\

% ...
$^{13}$Institute of Physics, Azerbaijan Academy of Sciences, Baku; Azerbaijan.\\
% ...
$^{40}$Niels Bohr Institute, University of Copenhagen, Copenhagen; Denmark.\\
% ...
$^{53}$II. Physikalisches Institut, Georg-August-Universit\"{a}t G\"ottingen, G\"ottingen; Germany.\\
% ...
$^{70}$$^{(a)}$INFN Sezione di Pavia;$^{(b)}$Dipartimento di Fisica, Universit\`a di Pavia, Pavia; Italy.\\
% ...
$^{93}$Department of Physics, Royal Holloway University of London, Egham; United Kingdom.\\
% ...
$^{101}$CPPM, Aix-Marseille Universit\'e, CNRS/IN2P3, Marseille; France.\\
$^{102}$Department of Physics, University of Massachusetts, Amherst MA; United States of America.\\
% ...
$^{128}$Homer L. Dodge Department of Physics and Astronomy, University of Oklahoma, Norman OK; United States of America.\\
% ...
$^{156}$Department of Physics and Astronomy, University of Sussex, Brighton; United Kingdom.\\
% ...
$^{167}$Department of Physics, University of Toronto, Toronto ON; Canada.\\
% ...
$^{a}$ Also at Centre for High Performance Computing, CSIR Campus, Rosebank, Cape Town; South Africa.\\
$^{b}$ Also at CERN, Geneva; Switzerland.\\
% ...
$^{*}$ Deceased

\end{flushleft}


\end{document}